\documentclass[preprint,floatfix] {revtex4} 
\newcommand{\rvec}{\mathrm {\mathbf {r}}} 
\newcommand{\pvec}{\mathrm {\mathbf {p}}} 
\usepackage{graphicx}
\usepackage{subfigure}
\usepackage{xcolor}
\usepackage{amsmath}
\usepackage{enumerate}
\usepackage{longtable}
\usepackage{amssymb}
\usepackage{tabularx}

\usepackage{color, soul}
\definecolor{darkblue}{rgb}{0,0,0.5}
\setulcolor{darkblue}
\begin{document}
\title{Some complexity measures in confined isotropic harmonic oscillator}

\author{Neetik Mukherjee}
%%\altaffiliation{Email: neetik.mukherjee@iiserkol.ac.in.}

\author{Amlan K.~Roy}
\altaffiliation{Corresponding author. Email: akroy@iiserkol.ac.in, akroy6k@gmail.com.}
\affiliation{Department of Chemical Sciences\\
Indian Institute of Science Education and Research (IISER) Kolkata, 
Mohanpur-741246, Nadia, WB, India}

\begin{abstract}
%%1234567890 %%1234567890 %%1234567890 %%1234567890 %%1234567890 %%1234567890 %%1234567890 %%1234567890 %%1234567890 %%1234567890
Various well-known statistical measures like \emph{L\'opez-Ruiz, Mancini, Calbet} (LMC) and \emph{Fisher-Shannon} complexity have 
been explored for confined isotropic harmonic oscillator (CHO) in composite position ($r$) and momentum ($p$) spaces. To get a 
deeper insight about CHO, a more generalized form of these quantities with R\'enyi entropy ($R$) is invoked here. The 
importance of scaling parameter in the exponential part is also investigated. $R$ is estimated considering order of entropic 
moments $\alpha, \beta$ as $(\frac{2}{3},3)$ in $r$ and $p$ spaces respectively. Explicit results of these measures with respect 
to variation of confinement radius $r_c$ is provided systematically for first eight energy states, 
namely, $1s,~1p,~1d,~2s,~1f,~2p,~1g$ and $2d$. Detailed analysis of these complexity measures provides 
many hitherto unreported interesting features. 
 
\vspace{5mm}

\vspace{5mm}

\end{abstract}
\maketitle
%123456789123456789123456789123456789123456789123456789123456789123456789123456789123456789123456789123456789123456789123456789123456789
\section{introduction}
Quantum particles undergo dramatic changes in their chemical and physical properties under extreme pressure.
Such situations may be achieved by shifting their spatial boundary from infinity to finite region \cite{michels37}. 
The effect of boundary condition and boundary limit on various observable properties such as energy spectrum, 
transition frequency, transition probability, polarizability, chemical reactivity, ionization potential etc., 
were studied in considerable detail in last ten years \cite{sabin2009,sen2014electronic}. These systems have their
comprehensive and potential application in nano-science and technology, condensed matter physics, semiconductor 
physics, quantum dot, quantum wells and quantum wires, etc \cite{sabin2009,sen2014electronic}. 

From the beginning of this century there has been a thriving interest in exploring statistical quantities namely, 
Fisher information ($I$), Onicescu energy ($E$), Shannon entropy ($S$) and R\'enyi entropy ($R$) as signifier of 
certain chemical, physical properties of a quantum system. In the same direction, \emph{complexity}, another topical 
concept, is directly concerned to aforesaid measures and illustrates their combined effect. A global definition of complexity
has not yet been possible. But it may be treated as a demonstrator of pattern, structure or correlation related with the 
distribution function in a given system. It depends on the scale of inspection, and comprises an important area of 
investigation with contemporary interest in chaotic systems, spatial patterns, language, multi-electronic systems, 
molecular or DNA analysis, social science, astrophysics and cosmology \cite{rosso03,shalizi04,chatzisavvas05,bouvrie11} etc.  

A quantum harmonic oscillator is a complex system; circumscribing its oscillation within an impenetrable region makes it even 
more impressive according to a complex world \cite{goldenfeld99,sen12}. Complexity, in a system, is introduced by disrupting 
certain rules of symmetry. A system possesses finite complexity when it is either in a state having less than some maximal 
order or not at a state of equilibrium. In a nutshell, it vanishes at two limiting cases, \emph{viz.}, when a system is 
(i) completely ordered (maximum distance from equilibrium) or (ii) at equilibrium (maximum disorder) \cite{sen12}. Overall 
it provides a characteristic idea of distribution in a system and is deliberated as a general descriptor of structure and 
correlation. In literature several definitions are available; some of them are Shiner, Davidson, Landsberg 
(\emph{SDL}) \cite{landsberg84,landsberg98,shiner99}, L\'opez-Ruiz, Mancini, Calbet (LMC) \emph{shape} ($C_{LMC}$) 
\cite{lopezruiz95,anteneodo96,catalan02,sanchez05}, \emph{Fisher-Shannon} ($C_{IS}$) \cite{romera04,sen07,angulo08}, 
\emph{Cram\'er-Rao} \cite{dehesa06,angulo08,antolin09} or \emph{Generalized R\'enyi-like} complexity 
\cite{calbet01,martin06,romera08,lopezruiz09}, generalized relative complexity measures \cite{romera11} etc.  

Without any loss of generality, the statistical measure of complexity, may be defined as a product of ordered and disordered 
parameters in the following form,
\begin{equation}
 C_{LMC} = H . D,
\end{equation}
where \emph{H} represents the information content and \emph{D} narrates an idea of concentration of spatial distribution. For 
a normalized continuous distribution $p(\rvec)$ these two quantities were expressed \cite{lopezruiz95} in the form 
$H=-k\int p(\rvec)\mathrm{log} \ p(\rvec)\mathrm{d}\rvec$ ($k$ is a positive constant) and $D=\int p^2(\rvec) \mathrm{d}\rvec$. 
But, this definition of $C_{LMC}$ was criticized \cite{feldman98} due to its inability to satisfy necessary conditions such as 
reaching minimal values for both extremely ordered and disordered limits, invariance under scaling, translation and replication. 
Therefore, this model was modified \cite{lopez05}, giving rise to the expression, 
\begin{equation}
C_{\emph{LMC}} = D . {e^S}.
\end{equation}
Here, $S$ quantifies the information of a given system and has the mathematical form $-\int p(\rvec)\mathrm{log} \ 
p(\rvec)\mathrm{d}\rvec$. Principally $C_{\emph{LMC}}$ quantifies the interaction between intrinsic information hidden in a 
system, and measure of a probabilistic distribution amongst its observed parts. It has potential application in several fields 
like detection of periodic, quasi-periodic, linear stochastic, chaotic dynamics \cite{lopezruiz95,yamano04,yamano04a} and in
quantum phase transition \cite{nagy12,romera14}. 

In information theory $E$ measures information content of a system. It sets off to minimum at equilibrium. Hence $E$ signifies a
descriptor of order. Whereas information entropies like $S, R$, become maximum at equilibrium, thereby implying disorder. 
Complexity quantifies the extent of countervail between order and disorder. In many instances, $E$ is replaced by $I$. So far 
in the literature $S$ has been primarily used as disorder parameter. $C_{IS}$ is another measure, attained by changing the 
pre-exponential global factor in $C_{LMC}$ by a local factor like $I$. It unites global and local characters while conserving 
the characteristics of complexity. Effectiveness of $C_{IS}$ can be reviewed by looking at numerous literature available for 
both \emph{free} and \emph{confined} atomic systems, including atomic shell structure, ionization process 
\cite{angulo08,sen07,antolin09,angulo08a,szabo08} etc. Recently a more generalized version was also designed that uses $R$ in place of 
$S$, in $C_{LMC}$ and $C_{IS}$ \cite{sen12}. Later, a scaling factor ($b$) was invoked in exponential part. 

About a decade ago, both $C_{LMC}$ and $C_{IS}$ were explored in the context of Bohr-like states of \emph{free} isotropic 
harmonic oscillator (IHO) in $r, p$ spaces \cite{sanudo08}. However, in \emph{confined} environment \emph{LMC} complexity 
has been investigated only for ground state of some model systems, like, CHO, confined h-atom, particle in spherical box and 
confined Helium atom \cite{romera09,nagy09}. Very recently, the present authors have pursued similar calculations for confined 
hydrogen atom  \cite{majumder17} and found that, parameter $b$ plays a key role in interpreting the property of a system. In 
this endeavor, our objective is to explore four different types of complexity emerging out of two order ($I, E$) and two 
disorder ($S, R$) parameters, in conjugate spaces, as functions of confinement radius ($r_c$). We take into account two $b$ 
values available in literature \cite{sen12}, and these are ($\frac{2}{3}$ for $C_{IS}$, 1 for $C_{LMC}$). All calculations 
were carried out using \emph{exact} wave functions of CHO in $r$ space. The $p$-space wave function is computed by applying 
numerical Fourier transform of $r$-space counterpart. In the end, representative calculation are done for eight low-lying states 
\emph{viz.,} $1s,~1p,~1d,~2s,~1f,~2p,~1g$ and $2d$. Presentation of the article is as follows. 
Section~II gives a brief outline of the theoretical method used; Sec.~III presents a thorough discussion on our results, while 
we conclude with a few remarks in Sec.~IV.

\section {Methodology}
The time-independent, non-relativistic wave function for a CHO system, in $r$ space can be written as, 
\begin{equation} 
\Psi_{n,l,m} (\rvec) = \psi_{n, l}(r)  \ Y_{l,m} (\Omega), 
\end{equation}
with $r$ and $\Omega$ representing radial distance and solid angle successively. Here $\psi_{n,l}(r)$ denote the radial part 
and $Y_{l,m}(\Omega)$ identifies spherical harmonics. The latter has following common form in both $r$ and $p$ spaces 
($P_{l}^{m} (\cos \theta)$ denotes usual associated Legendre polynomial), 
\begin{equation}
Y_{l,m} ({\Omega}) =\Theta_{l,m}(\theta) \ \Phi_m (\phi) = (-1)^{m} \sqrt{\frac{2l+1}{4\pi}\frac{(l-m)!}{(l+m)!}} 
\ P_{l}^{m}(\cos \theta)\ e^{-im \phi}.  
\end{equation} 

The relevant radial Schr\"odinger equation under the influence of confinement is, 
\begin{equation}
	\left[-\frac{1}{2} \ \frac{d^2}{dr^2} + \frac{l (l+1)} {2r^2} + v(r) +v_c (r) \right] \psi_{n,l}(r)=
	\mathcal{E}_{n,l}\ \psi_{n,l}(r),
\end{equation}
where $v(r)=\frac{1}{2}\omega^{2}r^{2}$ and $\omega$ is the oscillator frequency. Our required confinement effect is induced by 
invoking the following form of potential: $v_c(r) = +\infty$ for $r > r_c$, and $0$ for $r \leq r_c$, where $r_c$ denotes 
radius of confinement.

\emph{Exact} generalized radial wave function for a CHO is mathematically expressed as \cite{montgomery07}, 
\begin{equation}
\psi_{n_{r}, l}(r)= N_{n_{r}, l} \ r^{l} \ _{1}F_{1}\left[\frac{1}{2}\left(l+\frac{3}{2}-\frac{\mathcal{E}_{n_{r},l}}{\omega}\right),
(l+\frac{3}{2}),\omega r^{2}\right] e^{-\frac{\omega}{2}r^{2}}.
\end{equation}
Here, $N_{n, l}$ represents normalization constant and $\mathcal{E}_{n,l}$ corresponds to the energy of a given state distinguished by 
quantum numbers $n,l$, whereas $_1F_1\left[a,b,r\right]$ represents confluent hypergeometric 
function. Allowed energies are obtained by applying the boundary condition $\psi_{n,\ell} (0)= \psi_{n,\ell} \ (r_c)=0$ (except for 
$l=0$ states, only $\psi_{n,\ell} \ (r_c)=0$ ). In this 
work, generalized pseudospectral (GPS) method was employed to calculate $\mathcal{E}_{n,l}$ of these states. This method has 
provided very accurate results for various model and real systems including atoms, molecules, some of which could be found in the 
references \cite{roy04,sen06,roy15}. This is very well documented and therefore omitted here.

The $p$-space wave function is obtained from Fourier transform of $r$-space counterpart,  
\begin{equation}
\begin{aligned}
\psi_{n,l}(p) & = & \frac{1}{(2\pi)^{\frac{3}{2}}} \  \int_0^{r_{c}} \int_0^\pi \int_0^{2\pi} \psi_{n,l}(r) \ \Theta(\theta) 
 \Phi(\phi) \ e^{ipr \cos \theta}  r^2 \sin \theta \ \mathrm{d}r \mathrm{d} \theta \mathrm{d} \phi   \\
      & = & \frac{1}{2\pi} \sqrt{\frac{2l+1}{2}} \int_0^{r_{c}} \int_0^\pi \psi_{n,l} (r) \  P_{l}^{0}(\cos \theta) \ 
e^{ipr \cos \theta} \ r^2 \sin \theta  \ \mathrm{d}r \mathrm{d} \theta.  
\end{aligned}
\end{equation}
Here $\psi(p)$ is not normalized and needs to be normalized. Integrating over $\theta$ and $\phi$ yields,  
\begin{equation}
\psi_{n,l}(p)=(-i)^{l} \int_0^{r_{c}} \  \frac{\psi_{n,l}(r)}{p} \ f(r,p)\mathrm{d}r.    
\end{equation}
$f(r,p)$ depends only on $l$ quantum number. It can be expressed in terms of \emph{Cos} and \emph{Sin} series. More details about 
$f(r,p)$ could be found in \cite{mukherjee18}.
%Depending on $l$, this can be rewritten in following simplified form ($m$ starts with 0), 
%\begin{equation}
%\begin{aligned}
%f(r,p) & = & \sum_{k=2m+1}^{m<\frac{l}{2}} a_{k} \ \frac{\cos pr}{p^{k}r^{k-1}} +  
%\sum_{j=2m}^{m=\frac{l}{2}} b_{j} \ \frac{\sin pr}{p^{j}r^{j-1}}, \ \ \ \ \mathrm{for} \ \mathrm{even} \ l,   
%\\ f(r,p) & = & \sum_{k=2m}^{m=\frac{l-1}{2}} a_{k} \ \frac{\cos pr}{p^{k}r^{k-1}} +  
%\sum_{j=2m+1}^{m=\frac{l-1}{2}} b_{j} \ \frac{\sin pr}{p^{j}r^{j-1}}, \ \ \ \ \mathrm{for} \ \mathrm{odd} \ l.
%\end{aligned} 
%\end{equation}
%The values of coefficients $a_{k}$, $b_{j}$ of even-$l$ and odd-$l$ states can easily be computed from Eq.~(2).

The normalized position and momentum electron densities are expressed as,
\begin{equation}
\begin{aligned}
\rho(\rvec) = |\psi_{n,l,m}(\rvec)|^2 ,    \ \ \ \ \ \ 
\Pi(\pvec) = |\psi_{n,l,m} (\pvec)|^2 .
\end{aligned}
\end{equation}

Without any loss of generality, let us express complexity in a generalized mathematical form as $C = Ae^{b.B}$. The order ($A$) 
and disorder parameters ($B$) may comprise of ($E, I$) and ($R, S$) respectively. With this in mind, we are interested in the 
following four quantities, 
\begin{equation}
\begin{aligned}
C_{ER} & = E e^{bR}, \ \ \ \ \ \ \ \ C_{IR} = Ie^{bR}, \ \ \ \ \ \ \ C_{ES} & = E e^{bS}, \ \ \ \ \ \ \ \ C_{IS} = Ie^{bS}. 
\end{aligned} 
\end{equation}

Shannon entropy of a continuous density distribution is written as (`t' stands for \emph{total}), 
\begin{equation}
\begin{aligned} 
S_{\rvec} & =  -\int_{{\mathcal{R}}^3} \rho(\rvec) \ \ln [\rho(\rvec)] \ \mathrm{d} \rvec ;\ \ \ 
S_{\pvec} & =  -\int_{{\mathcal{R}}^3} \Pi(\pvec) \ \ln [\Pi(\pvec)] \ \mathrm{d} \pvec; \ \ \
S_{t} & = S_{\rvec}+S_{\pvec}. 
\end {aligned}
\end{equation}
Similarly, R{\'e}nyi entropy of order $\alpha (\neq 1)$ is obtained by taking logarithm of $\alpha$ and $\beta$-order 
entropic moments in respective spaces \cite{sen12}, 
\begin{equation}
\begin{aligned}
R_{\rvec}^{\alpha}  =  \frac{1}{1-\alpha} \ln \left(\int_{{\mathcal{R}}^3} \rho^{\alpha}(\rvec)\mathrm{d} \rvec \right) ;\ \ \ 
R_{\pvec}^{\beta}  =  \frac{1}{1-\beta} \ln \left[\int_{{\mathcal{R}}^3} \Pi^{\beta}(\pvec)\mathrm{d} \pvec \right]; \ \ \
R_{t} = R_{\rvec}^{\alpha}+R_{\pvec}^{\beta}.
\end {aligned}
\end{equation}
where,
\begin{equation}
\frac{1}{\alpha}+\frac{1}{\beta}=2 \nonumber 
\end{equation}
$I_{\rvec}$, $I_{\pvec}$ for a particle in a central potential may be expressed as \cite{romera05},
\begin{equation}
\begin{aligned} 
I_{\rvec} & =  4\langle p^2\rangle - 2(2l+1)|m|\langle r^{-2}\rangle;\ \ \
I_{\pvec} & =  4\langle r^2\rangle - 2(2l+1)|m|\langle p^{-2}\rangle; \ \ \
I_{t} = I_{\rvec}I_{\pvec}.  
\end{aligned} 
\end{equation}
Finally, $E$ is given by the following expressions in conjugate space \cite{sen12}, 
\begin{equation}
\begin{aligned}
E_{\rvec} & = \int_{{\mathcal{R}}^3}\rho^{2}(\rvec)\mathrm{d} \rvec;\ \ \ 
E_{\pvec} & = \int_{{\mathcal{R}}^3}\Pi^{2}(\pvec)\mathrm{d} \pvec;\ \ \
E_{t} = E_{\rvec}E_{\pvec}.  
\end{aligned}
\end{equation}

\begin{figure}                         %%%Fig. 1, CHA
\begin{minipage}[c]{0.38\textwidth}\centering
\includegraphics[scale=0.75]{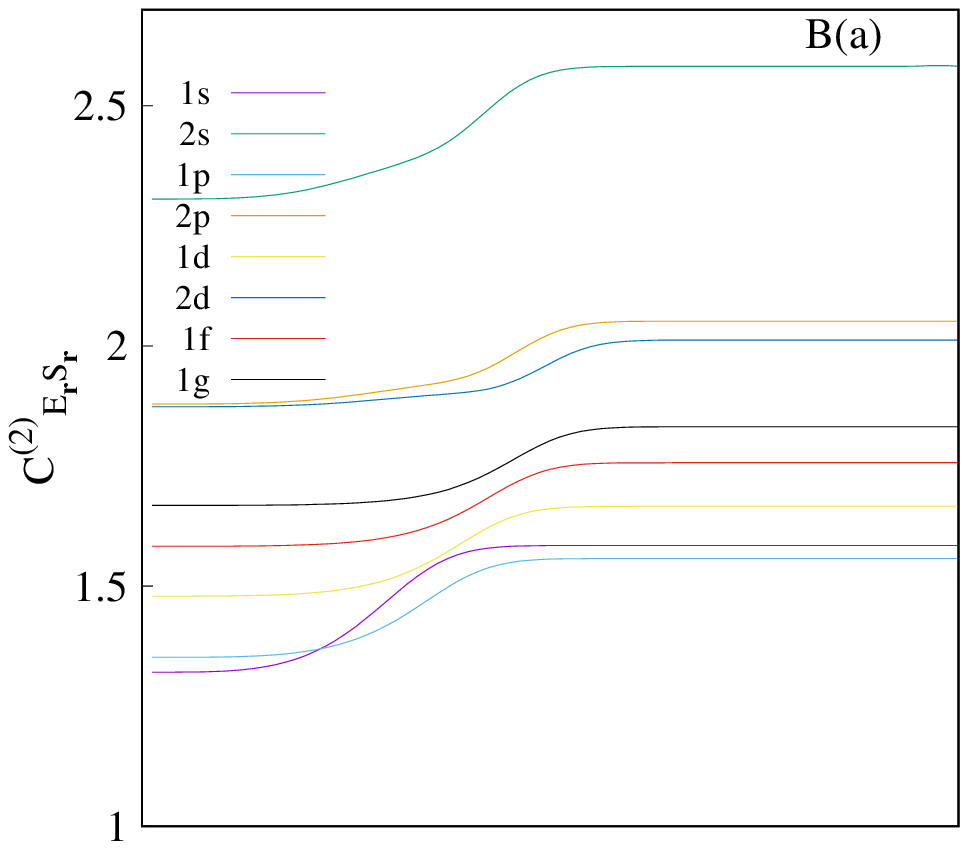}
\end{minipage}%
\hspace{10mm}
\begin{minipage}[c]{0.38\textwidth}\centering
\includegraphics[scale=0.75]{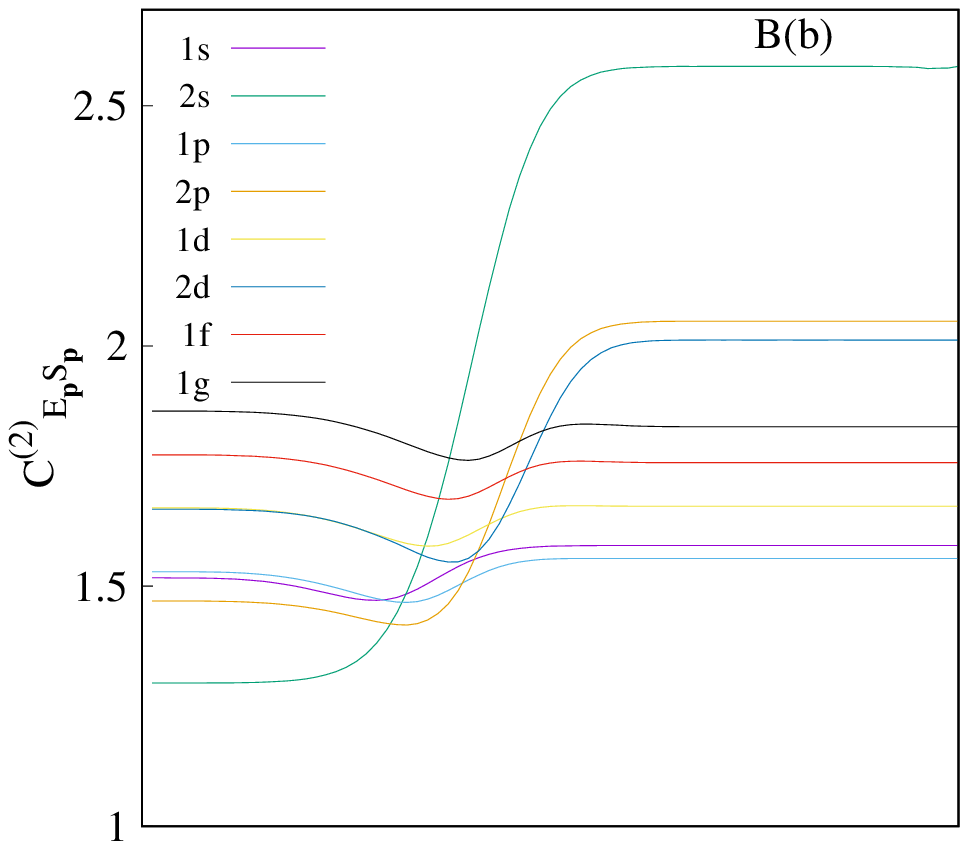}
\end{minipage}%
\vspace{5mm}
\hspace{0.5in}
\begin{minipage}[c]{0.39\textwidth}\centering
\includegraphics[scale=0.8]{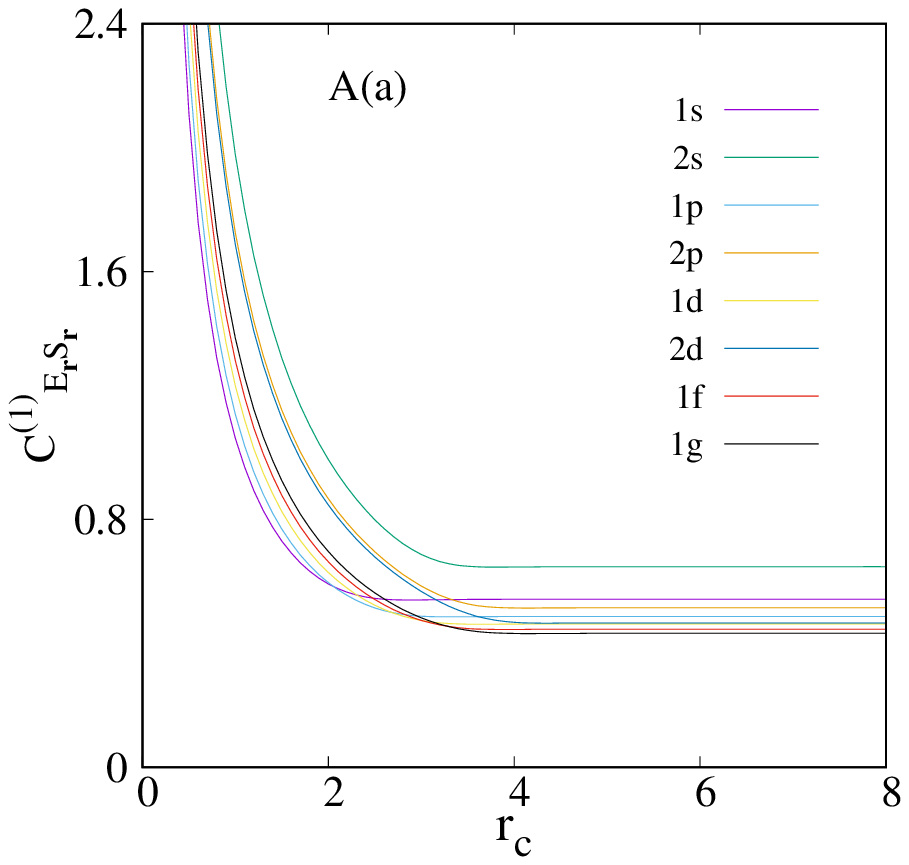}
\end{minipage}%
\hspace{10mm}
\begin{minipage}[c]{0.39\textwidth}\centering
\includegraphics[scale=0.8]{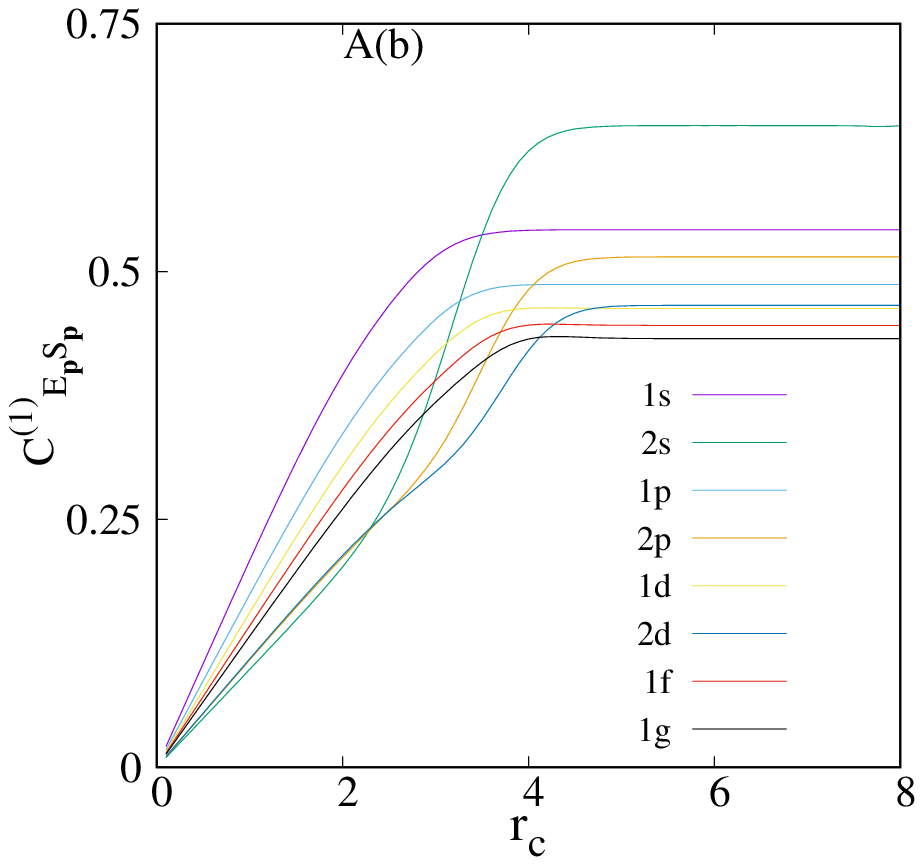}
\end{minipage}%
\caption{Variation of $C_{E_{\rvec}S_{\rvec}}^{(1)},~C_{E_{\pvec}S_{\pvec}}^{(1)}$ (bottom row A) and 
$C_{E_{\rvec}S_{\rvec}}^{(2)},~C_{E_{\pvec}S_{\pvec}}^{(2)}$ (top row B) in CHO with $r_c$ for $1s, 1p, 1d, 2s, 1f, 2p, 1g, 2d$
states.  See text for details.}
\end{figure}
 
\begingroup           %table1 
\squeezetable
\begin{table}
\caption{$C_{E_{\rvec}S_{\rvec}}^{(1)},~C_{E_{\pvec}S_{\pvec}}^{(1)}$ and $C_{E_{t}S_{t}}^{(2)}$ for $1s,~2s,~1p,~1d$ states in 
CHO at various $r_c$.}
\centering
\begin{ruledtabular}
\begin{tabular}{l|lll|lll}
$r_c$  &  $C_{E_{\rvec}S_{\rvec}}^{(1)}$ & $C_{E_{\pvec}S_{\pvec}}^{(1)}$  & $C_{E_{t}S_{t}}^{(1)}$\hspace{5mm} & 
$C_{E_{\rvec}S_{\rvec}}^{(1)}$ & $C_{E_{\pvec}S_{\pvec}}^{(1)}$  & $C_{E_{t}S_{t}}^{(1)}$ \vspace{1mm}  \\ 
\hline
\multicolumn{4}{c}{$1s$} \vline &  \multicolumn{3}{c}{\hspace{-5mm}$2s$}    \\
\hline
 0.1   & 10.544267      & 0.02093 & 0.22073     & 19.769038 &  0.00985  & 0.19491 \\
 0.2   & 5.272192       & 0.04186 & 0.22074     & 9.8845272 &  0.01972  & 0.19499 \\
 0.5   & 2.1098387      & 0.10464 & 0.22079     & 3.9539382 &  0.04933  & 0.19504 \\
 0.8   & 1.3220786      & 0.16721 & 0.22106     & 2.4716592 &  0.07894  & 0.19513 \\
 1.0   & 1.0623420      & 0.20854 & 0.22154     & 1.9779193 &  0.09871  & 0.19525 \\
 2.5   & 0.5453416      & 0.46594 & 0.25409     & 0.7991488 &  0.27391  & 0.21889 \\
 5.0   & 0.5422182      & 0.54221 & 0.29399     & 0.6472847 &  0.64698  & 0.41878 \\
 7.0   & 0.5422182865  & 0.5422182865 & 0.2940006701 & 0.6472880855    & 0.6472825716	& 0.4189782966 \\
\hline
\multicolumn{4}{c}{$1p$}  \vline  &   \multicolumn{3}{c}{\hspace{-5mm}$1d$}    \\
\hline
0.1  & 11.363226    & 0.01746    & 0.19849      & 12.302938    & 0.01561   & 0.19216  \\
0.2  & 5.6816369    & 0.03493    & 0.19850      & 6.1514809    & 0.03123   & 0.19217  \\
0.5  & 2.2730425    & 0.08733    & 0.19850      & 2.4607796    & 0.07809   & 0.19216  \\
0.8  & 1.4220367    & 0.13961    & 0.19853      & 1.5386569    & 0.12487   & 0.19213  \\
1.0  & 1.1395261    & 0.17426    & 0.19857      & 1.2318435    & 0.15593   & 0.19208  \\
2.5  & 0.5143143    & 0.40130    & 0.20639      & 0.5232749    & 0.36665   & 0.19186  \\
5.0  & 0.4869827    & 0.48699    & 0.23715      & 0.4628145    & 0.46285   & 0.21421  \\
7.0  & 0.4869829166 & 0.4869829186& 0.237152362 & 0.4628158714	& 0.4628158704	& 0.2141985303 \\
\end{tabular}
\end{ruledtabular}
\end{table}
\endgroup

\section {Result and Discussion}
At the onset it is convenient to point out a few points about this report. The \emph{net} information measures in conjugate 
$r$ and $p$ space may be separated into radial and angular segments. In a given space, the results provided correspond to 
\emph{net} measures including the \emph{angular} contributions. One can transform the IHO to a CHO by pressing the radial 
boundary of former from infinity to a finite region. This change in radial environment does not affect the \emph{angular} 
boundary conditions. Therefore, angular part of the information measures in unconfined and confined systems remain unaltered in 
both spaces. Further as we are solely focused in \emph{radial} confinement, this will also not influence the characteristics of 
a given measure as one changes $r_c$. Throughout this report, magnetic quantum number $m$ remains fixed to $0$. All the 
discussed measures of Eq.~(10) have been explored with change of $r_c$, choosing two different and widely used values of $b$ 
($1,\frac{2}{3}$). Note that, for $b=1$, $C_{ES}^{(2)}$ modifies to $C_{LMC}$; similarly $C_{IS}^{(1)}$ coincides with $C_{IS}$ 
at $b=\frac{2}{3}$. In order to simplify the discussion, a few words may be devoted to the notation followed. A uniform symbol 
$C_{order_{s}, disorder_{s}}^b$ is used; where the two subscripts refer to two order ($E, I$) and disorder ($S, R$) parameters. 
Another subscript $s$ is used to specify the space; \emph{viz.}, $r, p$ or $t$ (total). Two scaling parameters $b=\frac{2}{3}, 1$ 
are identified with superscripts (1), (2). These measures are offered systematically for $1s,~1p,~1d,~2s,~1f,~2p,~1g$ and $2d$ 
states in conjugate spaces, with $r_c$ varying in the range of 0.1-8.0 a.u. Also it is important to point out that, here levels 
are denoted by $n_r+1$ and $l$ values \cite{roy08}. Therefore, as an example, $n_r=0, l=1$ signify $1p$ state, and $n_r$ relates 
to $n$ as $n=2n_r+l$.  

\begin{figure}                         %%%Fig. 2, CHA
\begin{minipage}[c]{0.37\textwidth}\centering
\includegraphics[scale=0.75]{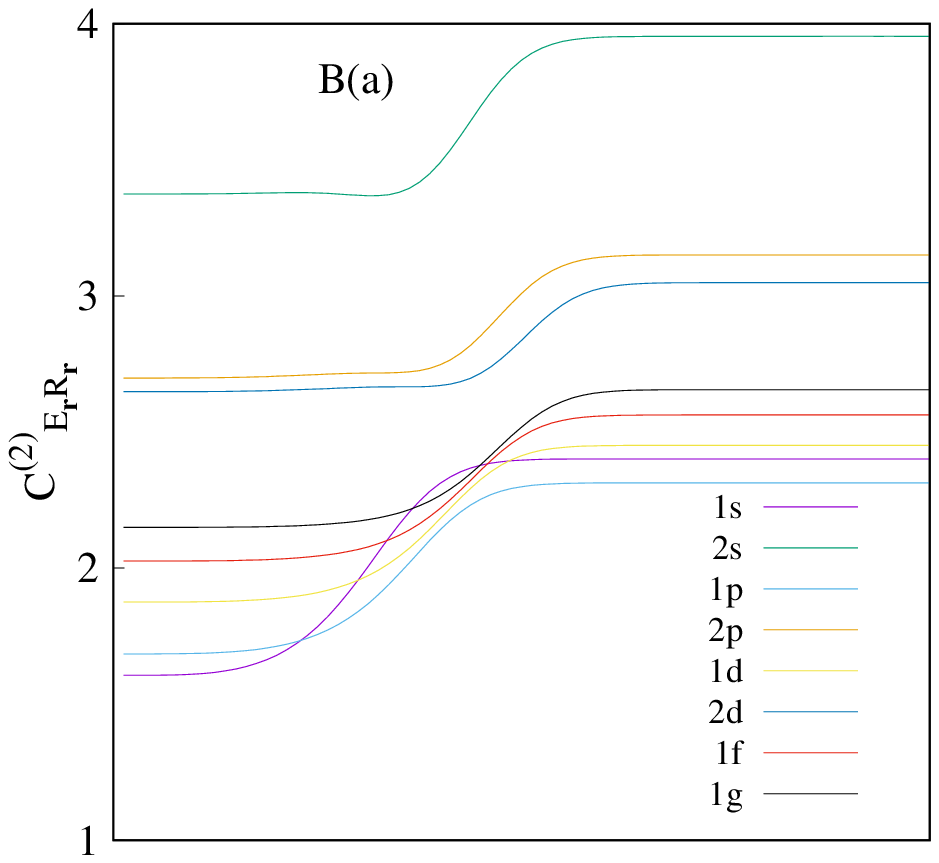}
\end{minipage}%
\hspace{10mm}
\begin{minipage}[c]{0.37\textwidth}\centering
\includegraphics[scale=0.75]{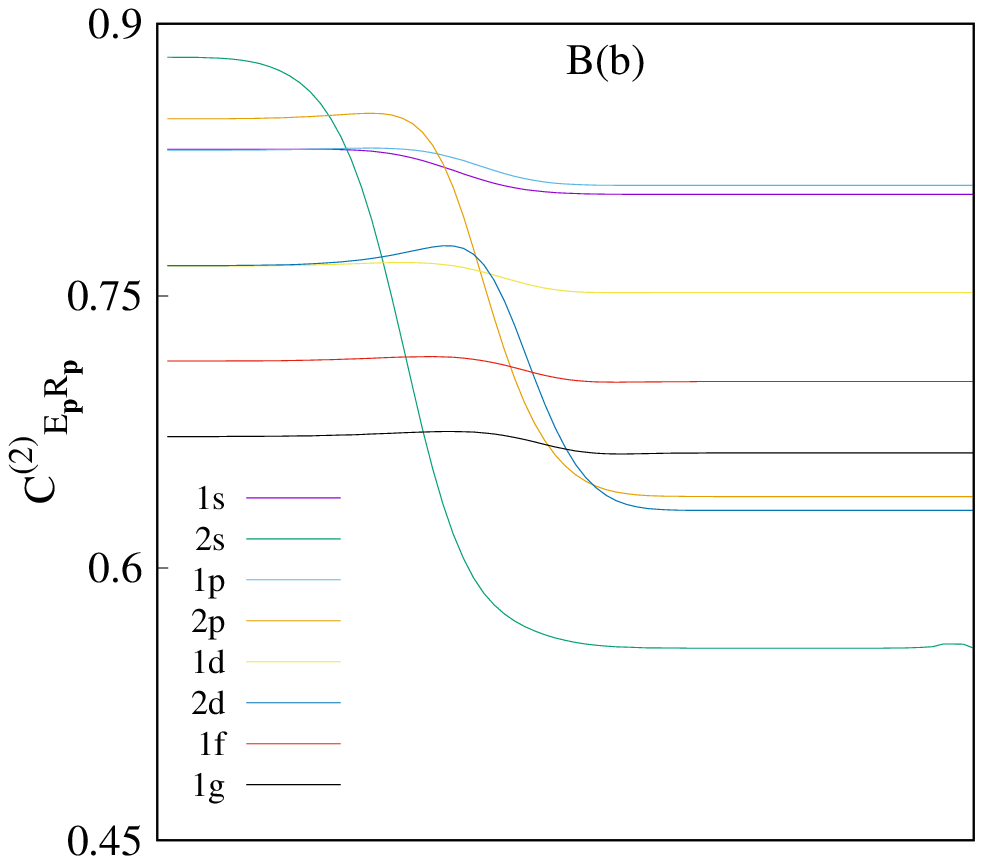}
\end{minipage}%
\vspace{5mm}
\hspace{0.5in}
\begin{minipage}[c]{0.39\textwidth}\centering
\includegraphics[scale=0.8]{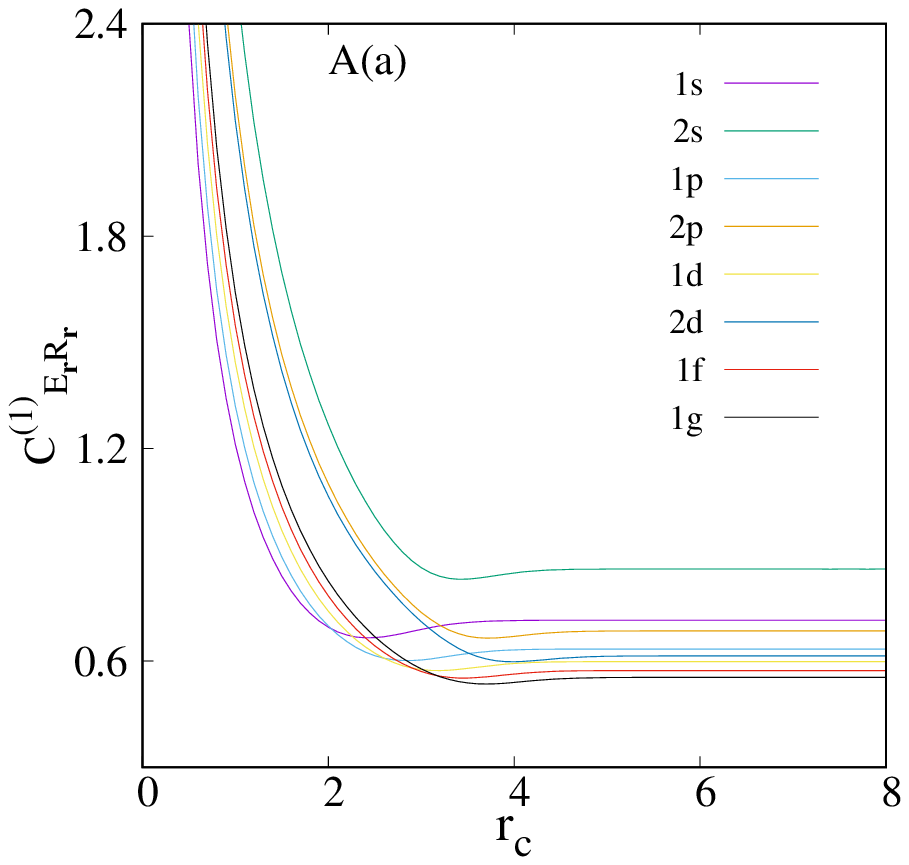}
\end{minipage}%
\hspace{10mm}
\begin{minipage}[c]{0.39\textwidth}\centering
\includegraphics[scale=0.8]{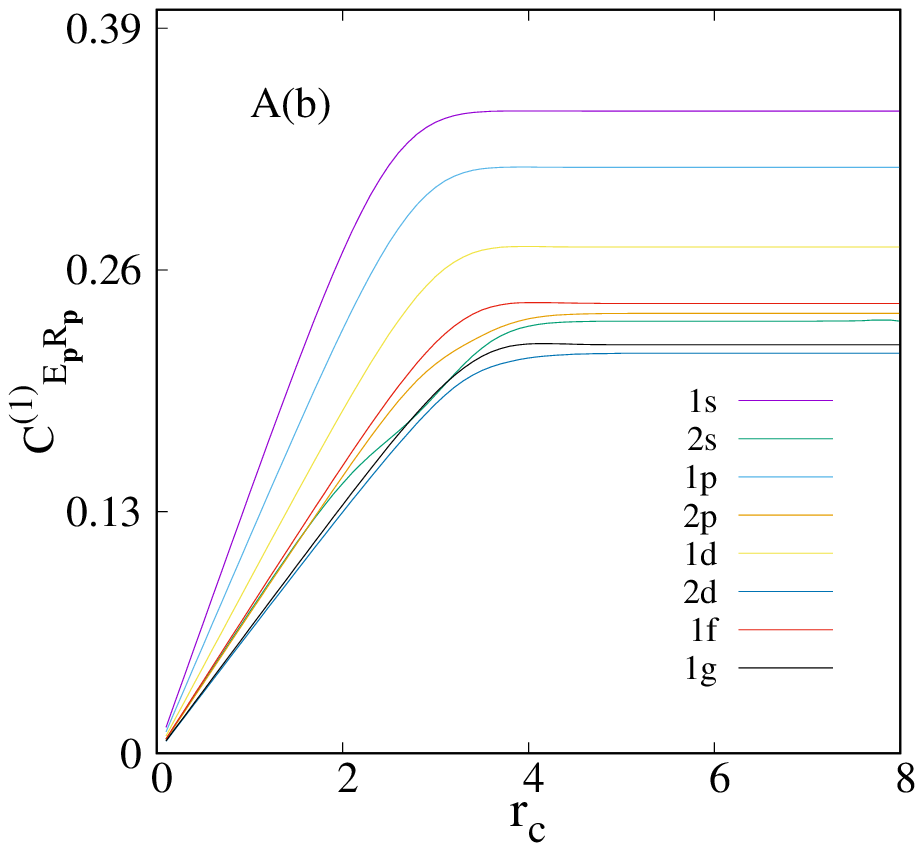}
\end{minipage}%
\caption{Changes in $C_{E_{\rvec}R_{\rvec}}^{(1)},~C_{E_{\pvec}R_{\pvec}}^{(1)}$ (bottom row A) and 
$C_{E_{\rvec}R_{\rvec}}^{(2)},~C_{E_{\pvec}R_{\pvec}}^{(2)}$ (top row B) in CHO with $r_c$ for $1s, 1p, 1d, 2s, 1f, 2p, 1g, 2d$ 
states. For more details, see text.}
\end{figure}

\begingroup           %table2
\squeezetable
\begin{table}
\caption{$C_{E_{\rvec}R_{\rvec}}^{(1)},~C_{E_{\pvec}R_{\pvec}}^{(1)}$ and $C_{E_{t}R_{t}}^{(1)}$ for $1s,~2s,~1p,~1d$ states in 
CHO at various $r_c$.}
\centering
\begin{ruledtabular}
\begin{tabular}{l|lll|lll}
$r_c$  &  $C_{E_{\rvec}R_{\rvec}}^{(1)}$ & $C_{E_{\pvec}R_{\pvec}}^{(1)}$  & $C_{E_{t}R_{t}}^{(1)}$\hspace{5mm} & 
$C_{E_{\rvec}R_{\rvec}}^{(1)}$ & $C_{E_{\pvec}R_{\pvec}}^{(1)}$  & $C_{E_{t}R_{t}}^{(1)}$ \vspace{1mm}  \\ 
\hline
\multicolumn{4}{c}{$1s$} \vline &  \multicolumn{3}{c}{\hspace{-5mm}$2s$}    \\
\hline
 0.1   & 12.011965    & 0.01401    & 0.16830    & 25.481700  & 0.00761  & 0.19405   \\
 0.2   & 6.0060677    & 0.02802    & 0.16830    & 12.740847  & 0.01523  & 0.19415   \\
 0.5   & 2.4038079    & 0.07004    & 0.16838    & 5.0962894  & 0.03810  & 0.19417   \\
 0.8   & 1.5073076    & 0.11199    & 0.16881    & 3.1849974  & 0.06091  & 0.19401   \\
 1.0   & 1.2125775    & 0.13981    & 0.16953    & 2.5477258  & 0.07602  & 0.19369   \\
 2.5   & 0.6661575    & 0.31636    & 0.21074    & 1.0079698  & 0.16895  & 0.17030   \\
 5.0   & 0.7153170    & 0.34549    & 0.24713    & 0.8597424  & 0.23246  & 0.19985   \\
 7.0   & 0.7153219525 & 0.3454938391 & 0.2471393275 & 0.8599102013	& 0.2324933318	& 0.1999233878   \\
\hline
\multicolumn{4}{c}{$1p$}  \vline  &   \multicolumn{3}{c}{\hspace{-5mm}$1d$}    \\
\hline
0.1  & 13.160407    & 0.01162     & 0.15295    & 14.410589   & 0.00931     & 0.13429    \\
0.2  & 6.5802381    & 0.02324     & 0.15296    & 7.2053113   & 0.01863     & 0.13430    \\
0.5  & 2.6326593    & 0.05811     & 0.15298    & 2.8823989   & 0.04659     & 0.13431    \\
0.8  & 1.6474274    & 0.09294     & 0.15312    & 1.8024800   & 0.07454     & 0.13435    \\
1.0  & 1.3207023    & 0.11610     & 0.15334    & 1.4433285   & 0.09314     & 0.13443    \\
2.5  & 0.6151214    & 0.27465     & 0.16894    & 0.6225095   & 0.22540     & 0.14031    \\
5.0  & 0.6339254    & 0.31520     & 0.19981    & 0.5985573   & 0.27230     & 0.16299    \\
7.0  & 0.6339446874 & 0.3152046743 & 0.1998223287 & 0.5986224459 & 0.2723045747	& 0.1630076305 \\
\end{tabular}
\end{ruledtabular}
\end{table}
\endgroup 

At first, in the bottom row (panels A(a)-A(b)) of Fig.~1, $C_{E_{\rvec}S_{\rvec}}^{(1)},~C_{E_{\pvec}S_{\pvec}}^{(1)}$ are
plotted against $r_c$ for all the eight states. Similarly, plots for $C_{E_{\rvec}S_{\rvec}}^{(2)},~C_{E_{\pvec}S_{\pvec}}^{(2)}$ 
are shown in panels B(a) and B(b) respectively. The lower panels clearly suggest that, $C_{E_{\rvec}S_{\rvec}}^{(1)}$ decreases 
and $C_{E_{\pvec}S_{\pvec}}^{(1)}$ increases with rise of $r_c$ before reaching a threshold corresponding to the IHO. The 
decrease of $C_{E_{\rvec}R_{\rvec}}^{(1)}$ with $r_c$ points to its inclination towards equilibrium. Next, panel A(a) reveals 
that at a fixed $l$, $C_{E_{\rvec}S_{\rvec}}^{(1)}$ progresses with $n$. Conversely, at a particular $n$, it reduces with growth 
of $l$. But panel A(b) does not imprint such patterns for $C_{E_{\pvec}S_{\pvec}}^{(1)}$. Panel A(c) in Figure~S1 of 
Supplementary Material (SM) presents that $C_{E_{t}S_{t}}^{(1)}$ enhances with advancement of $r_c$. Now, panels B(a), B(b) 
delineate the variation of $C_{E_{\rvec}S_{\rvec}}^{(2)},~C_{E_{\pvec}S_{\pvec}}^{(2)}$ with change of $r_c$. One sees that, 
$C_{E_{\rvec}S_{\rvec}}^{(2)}$ advances with growth of $r_c$ indicating that this is more prone towards order. On the other hand, 
at first $C_{E_{\pvec}S_{\pvec}}^{(2)}$ decreases with $r_c$, attains a minimum and finally coalesces to IHO. The ordering of 
$C_{E_{\rvec}S_{\rvec}}^{(2)}$ regarding $n,~l$ quantum numbers is akin to $C_{E_{\rvec}S_{\rvec}}^{(1)}$. It is noticed that, 
after a certain $r_c$ ($\gtrapprox 3$), both $C_{E_{\rvec}S_{\rvec}}^{(2)},~C_{E_{\pvec}S_{\pvec}}^{(2)}$ show analogous nature 
(increase to reach their respective limiting value). Panel B(c) in Fig.~S1 in SM portrays the increase of $C_{E_{t}S_{t}}^{(2)}$ 
with $r_c$. By comparing these two sets of complexity measure, namely $C^{(1)}_{ES}$ (in A(a)-A(c)) and $C^{(2)}_{ES}$ 
(in B(a)-B(c)), it is evident that, $C_{E_{\rvec}S_{\rvec}}^{(1)}$ and $~C_{E_{\pvec}S_{\pvec}}^{(1)}$ complement each other 
better (as former decreases and later increases with $r_c$) than the other set. Hence, we have presented 
$C_{E_{\rvec}S_{\rvec}}^{(1)},~C_{E_{\pvec}S_{\pvec}}^{(1)}$ and $C_{E_{t}S_{t}}^{(1)}$ at some selected $r_c$ values in Table~I, 
for $1s,~2s,~1p,~1d$. While for $2p,~2d,~1f,~1g$ states these are produced in Table~S1 of SM. These data consolidate the 
inference drawn from Figs.~1 and S1. We also see that when CHO approaches to IHO, $C_{E_{\rvec}S_{\rvec}}^{(1)}$ becomes equal 
to $~C_{E_{\pvec}S_{\pvec}}^{(1)}$. None of these results could be directly compared with literature data, as no such works have 
been published before, to the best of our knowledge.       

\begin{figure}                         %%%Fig. 3, CHA
\begin{minipage}[c]{0.38\textwidth}\centering
\includegraphics[scale=0.75]{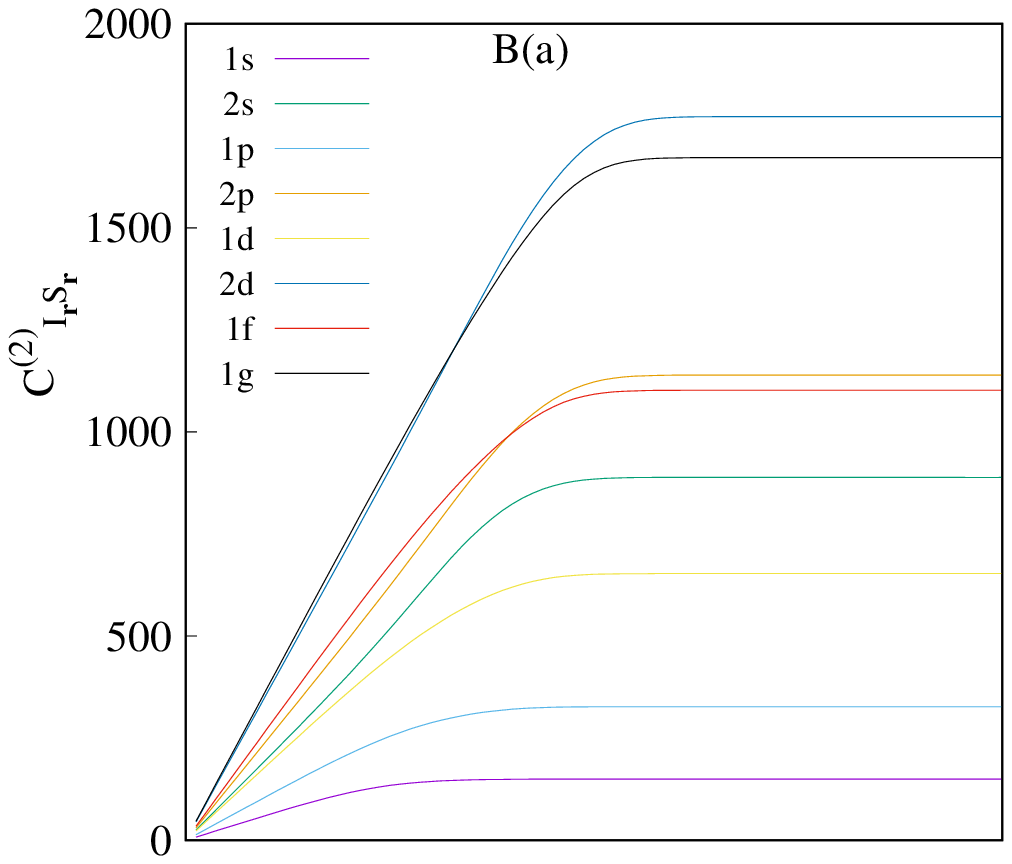}
\end{minipage}%
\hspace{15mm}
\begin{minipage}[c]{0.38\textwidth}\centering
\includegraphics[scale=0.75]{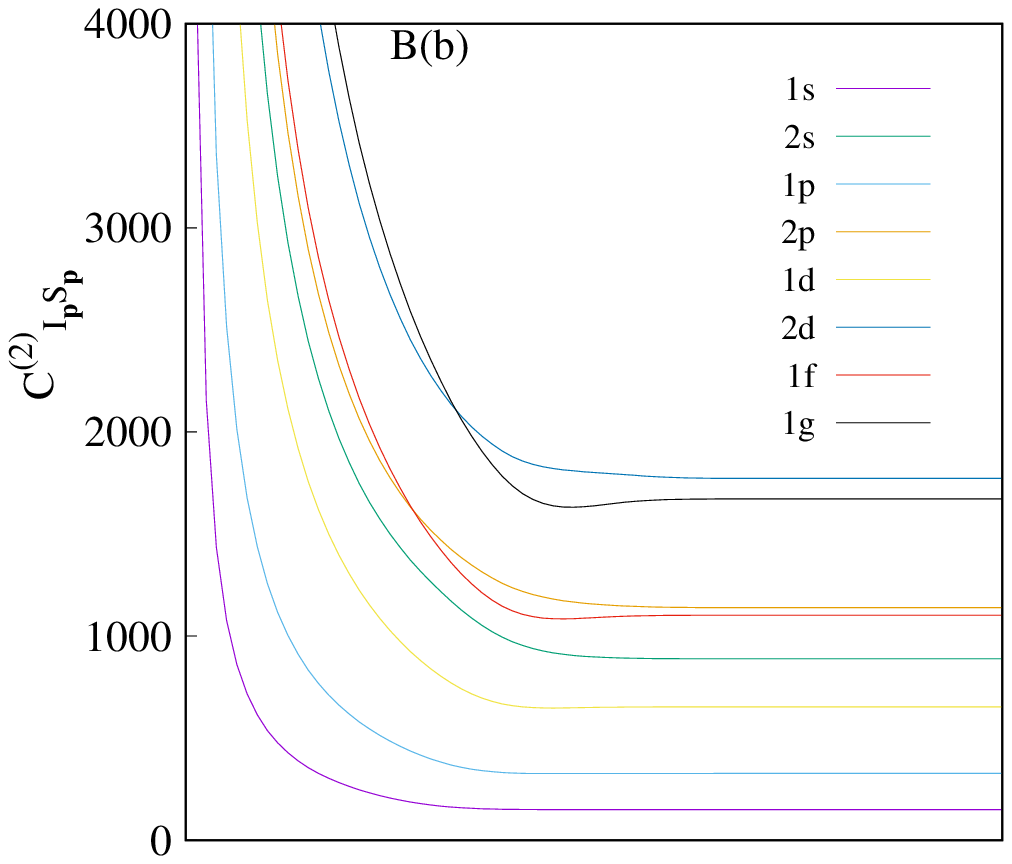}
\end{minipage}%
\vspace{5mm}
\hspace{0.6in}
\begin{minipage}[c]{0.39\textwidth}\centering
\includegraphics[scale=0.8]{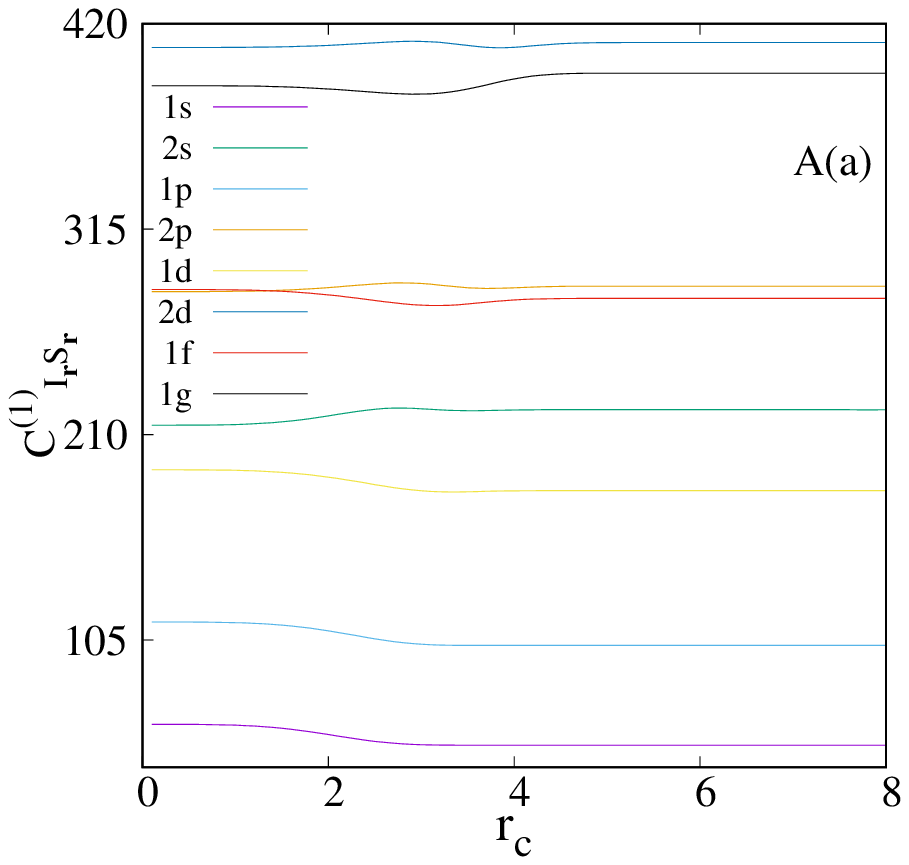}
\end{minipage}%
\hspace{12mm}
\begin{minipage}[c]{0.39\textwidth}\centering
\includegraphics[scale=0.8]{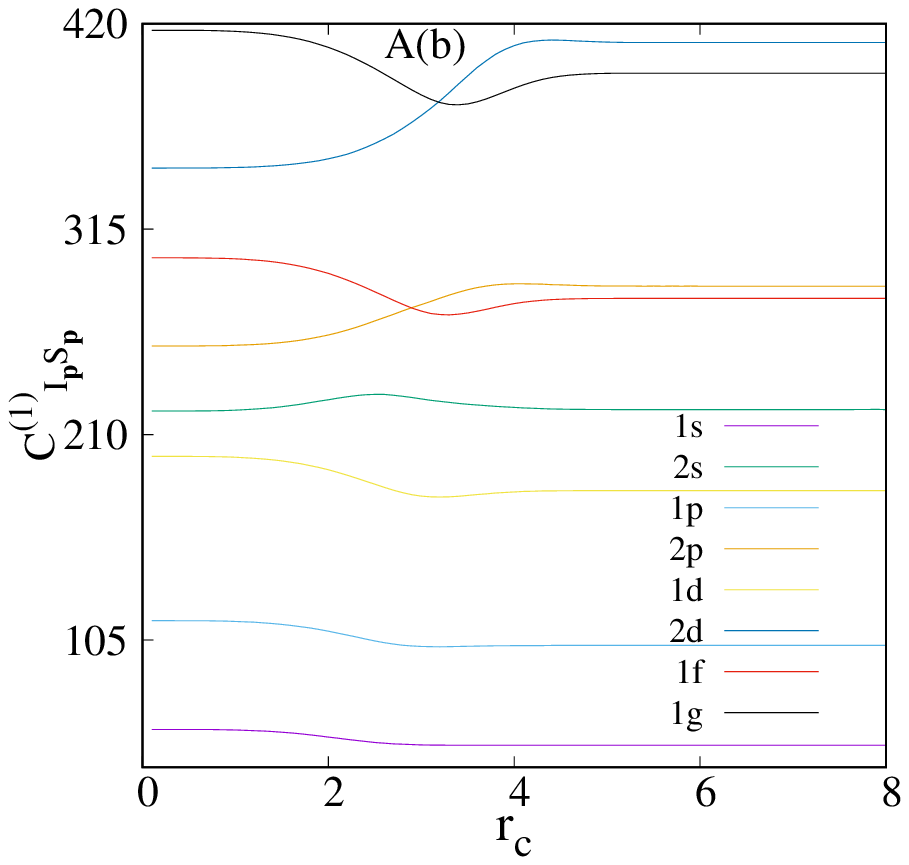}
\end{minipage}%
\caption{Variation of $C_{I_{\rvec}S_{\rvec}}^{(1)},~C_{I_{\pvec}S_{\pvec}}^{(1)}$ (bottom row A)  
$C_{I_{\rvec}S_{\rvec}}^{(2)},~C_{I_{\pvec}S_{\pvec}}^{(2)}$ (top row B)  
in CHO with $r_c$ for $1s, 1p, 1d, 2s, 1f, 2p, 1g, 2d$ states. Consult text for more details.} 
\end{figure}

\begingroup           %table3
\squeezetable
\begin{table}
\caption{$C_{I_{\rvec}S_{\rvec}}^{(2)},~C_{I_{\pvec}S_{\pvec}}^{(2)}$ and $C_{I_{t}S_{t}}^{(2)}$ for $1s,~2s,~1p,~1d$ states in CHO at various $r_c$.}
\centering
\begin{ruledtabular}
\begin{tabular}{l|lll|lll}
$r_c$  &  $C_{I_{\rvec}S_{\rvec}}^{(2)}$ & $C_{I_{\pvec}S_{\pvec}}^{(2)}$  & $C_{I_{t}S_{t}}^{(2)}$\hspace{5mm} & 
$C_{I_{\rvec}S_{\rvec}}^{(2)}$ & $C_{I_{\pvec}S_{\pvec}}^{(2)}$  & $C_{I_{t}S_{t}}^{(2)}$ \vspace{1mm}  \\ 
\hline
\multicolumn{4}{c}{$1s$} \vline &  \multicolumn{3}{c}{\hspace{-5mm}$2s$}    \\
\hline
 0.1   & 7.758205         & 4303.15991   & 33384.80018    & 25.056165  & 29216.5721 & 732055.2573 \\
 0.2   & 15.516171        & 2151.56736   & 33384.08737    & 50.112519  & 14608.3337 & 732060.4179 \\
 0.5   & 38.766040        & 860.42282    & 33355.18588    & 125.300544 & 5844.1079  & 732269.9114 \\
 0.8   & 61.804468        & 537.04703    & 33191.90621    & 200.655774 & 3655.3318  & 733463.4393 \\
 1.0   & 76.791106        & 428.68905    & 32919.50741    & 251.188685 & 2928.0196  & 735485.4165 \\
 2.5   & 145.231258       & 167.01717    & 24256.11393    & 663.377452 & 1216.9057  & 807267.8370 \\
 5.0   & 149.733084       & 149.73309    & 22419.99774    & 888.731009 & 888.9964   & 790078.7475 \\
 7.0   & 149.733087619267 & 149.73308721 & 22419.99746802 & 888.738383 & 888.7381   & 789855.7135 \\
\hline
\multicolumn{4}{c}{$1p$}  \vline  &   \multicolumn{3}{c}{\hspace{-5mm}$1d$}    \\
\hline
0.1  & 13.589616     & 10065.70616 & 136789.08528   & 23.086162        & 21178.15365 & 488922.29249    \\
0.2  & 27.179055     & 5032.83116  & 136787.59609   & 46.172182        & 10589.04165 & 488919.15916    \\
0.5  & 67.929622     & 2012.77692  & 136727.17589   & 115.415992       & 4235.04577  & 488792.00959    \\
0.8  & 108.523492    & 1256.72638  & 136384.33569   & 184.533851       & 2644.87901  & 488069.71221    \\
1.0  & 135.307764    & 1003.69056  & 135807.12614   & 230.387985       & 2113.17739  & 486850.68272    \\
2.5  & 293.849512    & 382.32570   & 112346.22278   & 535.936973       & 802.82774   & 430265.07166    \\
5.0  & 327.026856    & 327.02367   & 106945.52430   & 653.074539       & 653.04527   & 426487.24482    \\
7.0  & 327.027007440 & 327.0270080 & 106946.6637782 & 653.077279415428 & 653.077279  & 426509.932617    \\
\end{tabular}
\end{ruledtabular}
\end{table}
\endgroup           

Similarly, bottom row of Fig.~2 interprets the behavior of $C_{E_{\rvec}R_{\rvec}}^{(1)},~C_{E_{\pvec}R_{\pvec}}^{(1)}$ with 
$r_{c}$ for the same states of Fig.~1. Panel A(a) reveals that, $C_{E_{\rvec}R_{\rvec}}^{(1)}$ diminishes with growth of $r_c$, then 
attains a minimum and finally converges to respective IHO result. This minimum gets flatter with progress of both $n,~l$. Here also 
$C_{E_{\rvec}R_{\rvec}}^{(1)}$ shows analogous trend to what is noticed for $C_{E_{\rvec}S_{\rvec}}^{(1)}$, \emph{viz.}, (i) at a 
fixed $n$, both measures in $r$-space decline with growth of $l$ (ii) at a particular $l$, they accelerate with $n$ (iii) like 
$C_{E_{\rvec}S_{\rvec}}^{(1)}$, $C_{E_{\rvec}R_{\rvec}}^{(1)}$ is more inclined towards disorder. From panel A(b) it is also vivid 
that, like $C_{E_{\pvec}S_{\pvec}}^{(1)}$, $C_{E_{\pvec}R_{\pvec}}^{(1)}$ progress with growth of $r_c$. Additionally, at a definite 
$l$, $C_{E_{\pvec}R_{\pvec}}^{(1)}$ falls off with $n$. At a given $n$, it also reduces as $l$ advances. The relevant total 
measures are displayed in panel A(c) of Fig.~S2 of SM, where prominent minima are seen for $s$ states. As usual,  
$C_{E_{t}R_{t}}^{(1)}$ finally merge to IHO case. Panels B(a), B(b) in top row of Fig.~2 exhibit variations of 
$C_{E_{\rvec}R_{\rvec}}^{(2)},~C_{E_{\pvec}R_{\pvec}}^{(2)}$ with $r_c$. At smaller $r_c$ region ($\lessapprox 3$), they both 
change very slowly. At around $r_c=3$, $C_{E_{\rvec}R_{\rvec}}^{(2)}$ jumps and $C_{E_{\pvec}R_{\pvec}}^{(2)}$ drops to reach the 
IHO limit. For both $C_{E_{\rvec}R_{\rvec}}^{(2)}$ and $C_{E_{\pvec}R_{\pvec}}^{(2)}$ absolute values of the slope 
of the curve enhance as $n$ grows (fixed $l$) and decrease with growth of $l$ (fixed $n$). The dependence of 
$C_{E_{\rvec}R_{\rvec}}^{(2)}$ on $n,~l$ is similar to $C_{E_{\rvec}R_{\rvec}}^{(1)}$. Panel B(c) of Fig.~S2 in SM imprints the 
alteration of $C_{E_{t}R_{t}}^{(2)}$ with $r_c$ varying from 0-8. Once again one may conclude that, out of $C^{(1)}_{ER}$ and 
$C^{(2)}_{ER}$, the former offers a more clearer knowledge about CHO, which justifies the quantities produced in Table~II, namely, 
$C_{E_{\rvec}R_{\rvec}}^{(1)}$, $C_{E_{\pvec}R_{\pvec}}^{(1)}$ and $C_{E_{t}R_{t}}^{(1)}$. These are given for $1s,~1p,~2s,~1d$ 
states at eight suitably chosen $r_c$, whereas Table~S2 reports the same for $2p,~1f,~2d,~1g$ states. These results support the 
conclusions drawn from Figs.~2 and S2. Moreover it is also apparent that, there appears a minimum in $C_{E_{\rvec}R_{\rvec}}^{(1)}$ 
for all states. Similar to previous table, in this case also no literature values could be quoted. Additionally, 
$C_{E_{\pvec}R_{\pvec}}^{(1)}$ shows more lucid trend than $C_{E_{\pvec}S_{\pvec}}^{(1)}$ with respect to dependence of quantum 
numbers $n,~l$. Hence, in practice $C^{(1)}_{ER}$ may possibly be considered a better measure of complexity than $C^{(1)}_{ES}$.

\begin{figure}                         %%%Fig. 4, CHA
\begin{minipage}[c]{0.38\textwidth}\centering
\includegraphics[scale=0.75]{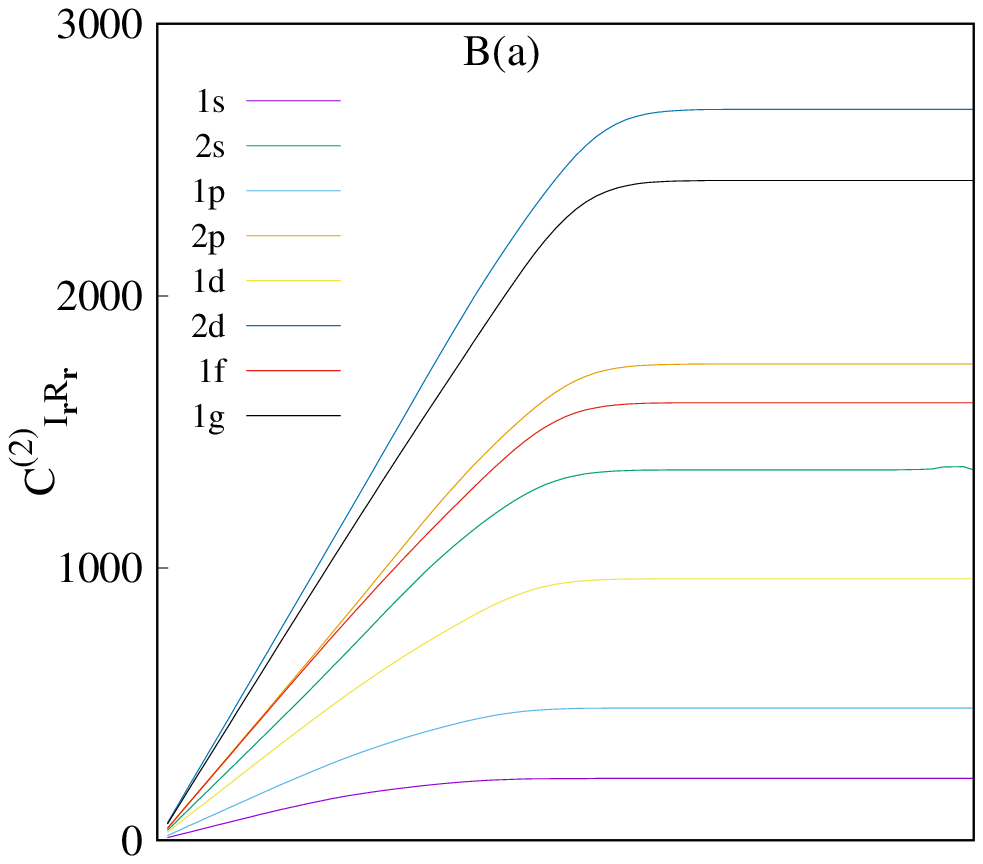}
\end{minipage}%
\hspace{10mm}
\begin{minipage}[c]{0.38\textwidth}\centering
\includegraphics[scale=0.75]{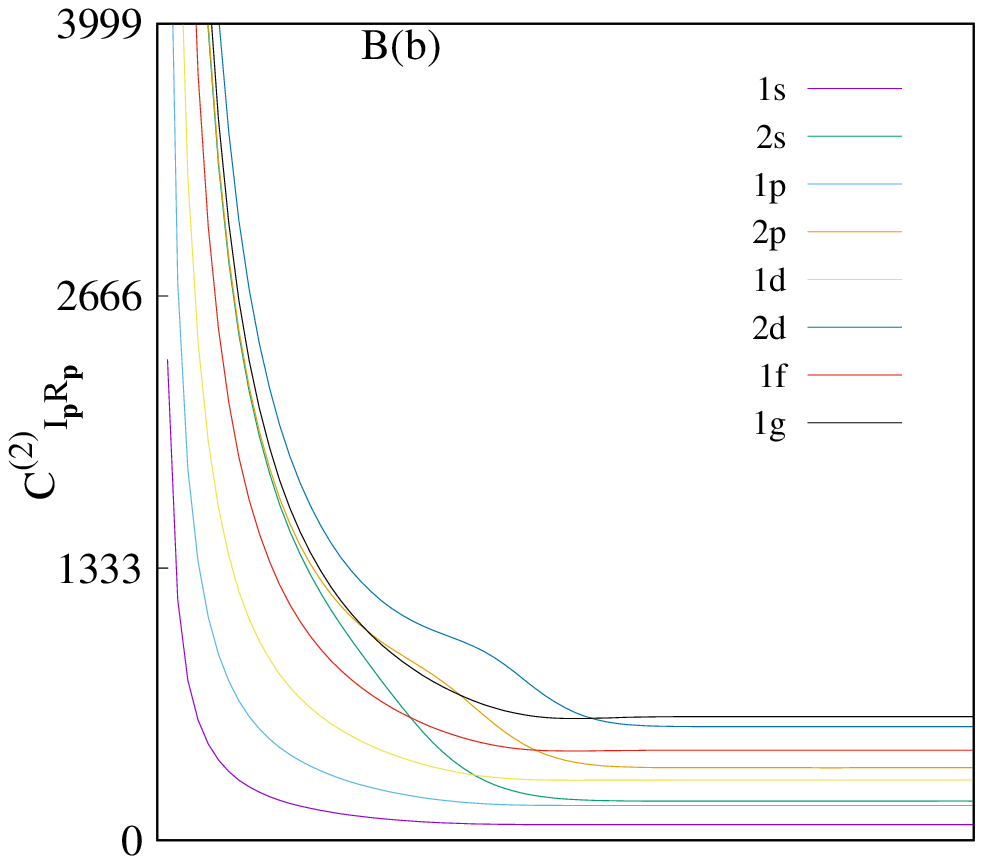}
\end{minipage}%
\vspace{5mm}
\hspace{0.5in}
\begin{minipage}[c]{0.39\textwidth}\centering
\includegraphics[scale=0.8]{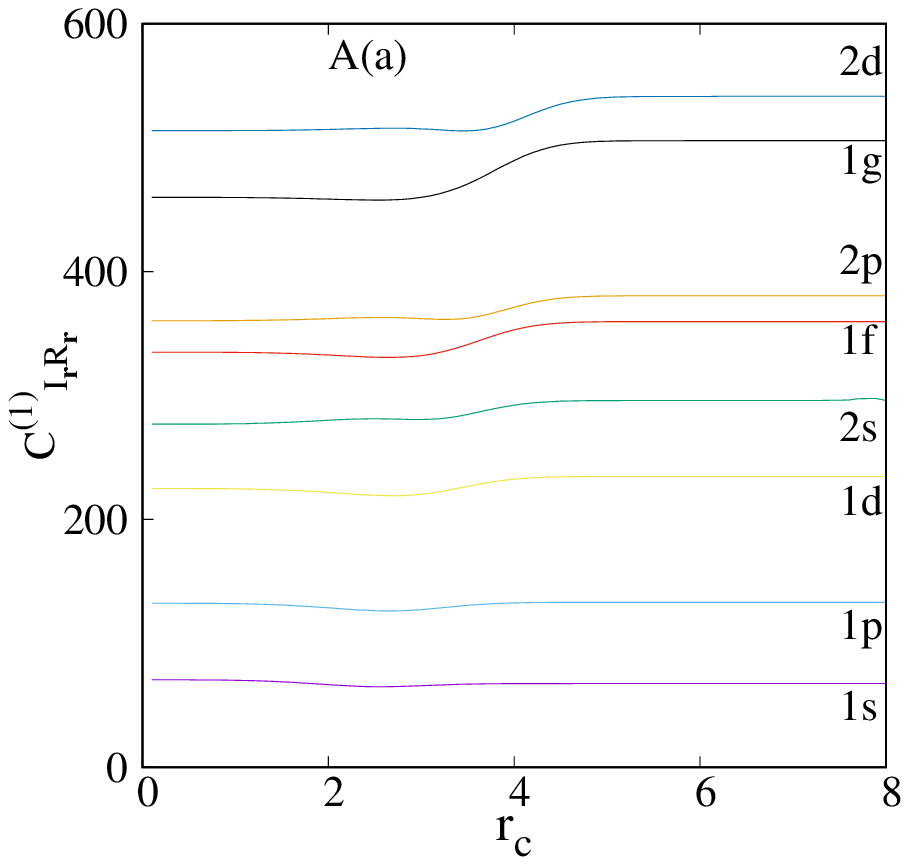}
\end{minipage}%
\hspace{10mm}
\begin{minipage}[c]{0.39\textwidth}\centering
\includegraphics[scale=0.8]{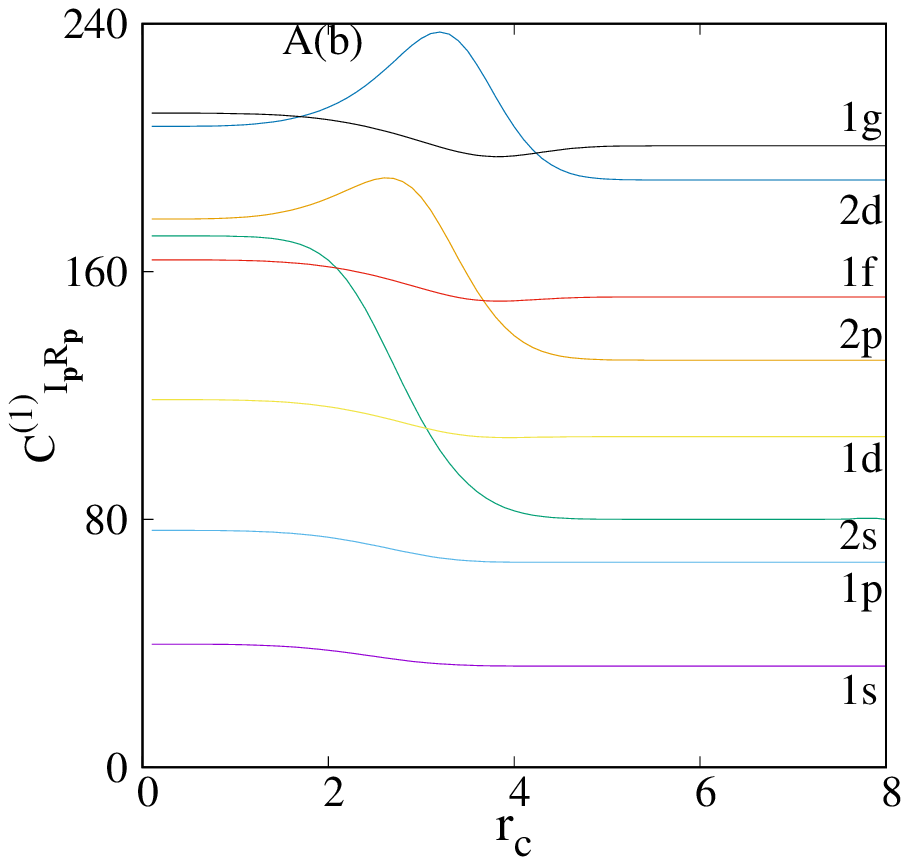}
\end{minipage}%
\caption{Plots of $C_{I_{\rvec}R_{\rvec}}^{(1)},~C_{I_{\pvec}R_{\pvec}}^{(1)}$ (bottom row A) and 
$C_{I_{\rvec}R_{\rvec}}^{(2)},~C_{I_{\pvec}R_{\pvec}}^{(2)}$ (top row B) in CHO with $r_c$ for $1s, 1p, 1d, 2s, 1f, 2p, 1g, 2d$ 
states. For further details, see text.}
\end{figure}

In Fig.~3, lower \{A(a), A(b)\} and upper \{B(a), B(b)\} panels depict the alteration of 
$\{C_{I_{\rvec}S_{\rvec}}^{(1)},~C_{I_{\pvec}S_{\pvec}}^{(1)}\}$ and 
$\{C_{I_{\rvec}S_{\rvec}}^{(2)},~C_{I_{\pvec}S_{\pvec}}^{(2)}\}$ with rise in $r_{c}$ for all the states mentioned above. Nature 
of variation of $\{C_{I_{\rvec}S_{\rvec}}^{(1)},~C_{I_{\pvec}S_{\pvec}}^{(1)}\}$ with $r_c$ changes from state to state. From lower 
panels it is gathered that, at a fixed $n$, both $C_{I_{\rvec}S_{\rvec}}^{(1)}$ and $C_{I_{\pvec}S_{\pvec}}^{(1)}$ elevate with $l$. 
Similarly from panel A(c) in Fig.~S3 of SM one can infer that, at a certain $n$, $C_{I_{t}S_{t}}^{(1)}$ increases with growth of 
$l$. On the other hand, the top panels B(a) and B(b) portray that, for these states $C_{I_{\rvec}S_{\rvec}}^{(2)}$ and 
$C_{I_{\pvec}S_{\pvec}}^{(2)}$ progress and regress with growth in $r_c$ and finally approach to respective IHO values. Like the 
previous cases, panel B(c) in Fig.~S3 shows the plot of $C_{I_{t}S_{t}}^{(2)}$ versus $r_c$. From panels \{B(a), B(b)\} one 
observes that, at a particular $n$ both $C_{I_{\rvec}S_{\rvec}}^{(2)}$, $C_{I_{\pvec}S_{\pvec}}^{(2)}$ increase with advancement 
in $l$. Also, at a certain $l$, they enhance with improvement of $n$. A careful study of Figs.~3 and S3 express that, in case of 
CHO, $\{C_{I_{\rvec}S_{\rvec}}^{(2)},~C_{I_{\pvec}S_{\pvec}}^{(2)},~C_{I_{t}S_{t}}^{(2)}\}$ offer more transparent pattern than 
$\{C_{I_{\rvec}S_{\rvec}}^{(1)},~C_{I_{\pvec}S_{\pvec}}^{(1)},~C_{I_{t}S_{t}}^{(1)}\}$. So, in order to get a quantitative idea,      
$\{C_{I_{\rvec}S_{\rvec}}^{(2)},~C_{I_{\pvec}S_{\pvec}}^{(2)},~C_{I_{t}S_{t}}^{(2)}\}$ values at eight different $r_c$'s are 
provided in Tables~III ($1s,~1p,~2s,~1d$) and S3 ($2p,~2d,~1f,~1g$). Again no results are available in literature.     

\begingroup           %table4
\squeezetable
\begin{table}
\caption{$C_{I_{\rvec}R_{\rvec}}^{(2)},~C_{I_{\pvec}R_{\pvec}}^{(2)}$ and $C_{I_{t}R_{t}}^{(2)}$ for $1s,~2s,~1p,~1d$ states in 
CHO at various $r_c$.}
\centering
\begin{ruledtabular}
\begin{tabular}{l|lll|lll}
$r_c$  &  $C_{I_{\rvec}R_{\rvec}}^{(2)}$ & $C_{I_{\pvec}R_{\pvec}}^{(2)}$  & $C_{I_{t}R_{t}}^{(2)}$\hspace{5mm} & 
$C_{I_{\rvec}R_{\rvec}}^{(2)}$ & $C_{I_{\pvec}R_{\pvec}}^{(2)}$  & $C_{I_{t}R_{t}}^{(2)}$ \vspace{1mm}  \\ 
\hline
\multicolumn{4}{c}{$1s$} \vline &  \multicolumn{3}{c}{\hspace{-5mm}$2s$}    \\
\hline
 0.1   & 9.433173     & 2356.25770  & 22226.98694   & 36.667294   & 19837.6528 & 727393.0643   \\
 0.2   & 18.866136    & 1178.12662  & 22226.69794   & 73.334753   & 9918.7949  & 727392.3819   \\
 0.5   & 47.144086    & 471.21450   & 22214.97745   & 183.353522  & 3967.0032  & 727364.0186   \\
 0.8   & 75.237916    & 294.38160   & 22148.65828   & 293.516420  & 2477.4726  & 727178.9180   \\
 1.0   & 93.643707    & 235.33582   & 22037.71925   & 367.211469  & 1979.1660  & 726772.4766   \\
 2.5   & 196.075111   & 93.44035    & 18321.32860   & 939.702768  & 589.5390   & 553991.4457   \\
 5.0   & 226.884302   & 76.15910    & 17279.30538   & 1360.441254 & 191.4695   & 260483.1381   \\
 7.0   & 226.88661187 & 76.1582577  & 17279.28906   & 1360.840385 & 191.307601 & 260358.110977 \\
\hline
\multicolumn{4}{c}{$1p$}  \vline  &   \multicolumn{3}{c}{\hspace{-5mm}$1d$}    \\
\hline
0.1  & 16.937866    & 5462.96222  & 92530.92231   & 29.265843      & 9760.87960  & 285660.37271    \\
0.2  & 33.875565    & 2731.47795  & 92530.35998   & 58.531547      & 4880.43593  & 285659.46552    \\
0.5  & 84.672004    & 1092.53992  & 92507.54481   & 146.314712     & 1952.11162  & 285622.65198    \\
0.8  & 135.321527   & 682.65619   & 92378.07852   & 233.974746     & 1219.84752  & 285413.51636    \\
1.0  & 168.827692   & 545.88248   & 92160.08039   & 292.195994     & 975.57961   & 285060.45712    \\
2.5  & 384.346677   & 216.47518   & 83201.51983   & 695.405048     & 386.96541   & 269097.70008    \\
5.0  & 485.702587   & 170.28809   & 82709.36712   & 960.529605     & 294.68695   & 283055.54874    \\
7.0  & 485.72457911 & 170.2950413 & 82716.48726   & 960.6865379473 & 294.7369698 & 283149.8391222    \\
\end{tabular}
\end{ruledtabular}
\end{table}
\endgroup  

Finally, in Fig.~4 the bottom (A(a),A(b)) and top (B(a)-B(b)) panels provide the behavioral pattern in our last complexity 
measure, \emph{viz.,} $C_{I_{\rvec}R_{\rvec}}^{(1)},~C_{I_{\pvec}R_{\pvec}}^{(1)}$ and 
$C_{I_{\rvec}R_{\rvec}}^{(2)},~C_{I_{\pvec}R_{\pvec}}^{(2)}$ with variations in $r_{c}$. Panel A(a) reveals that, 
$C_{I_{\rvec}R_{\rvec}}^{(1)}$ progresses slowly with $r_c$ to attain the IHO values. At a suitable $n$ this quantity advances 
with $l$. In a parallel manner, at a constant $l$, $C_{I_{\rvec}R_{\rvec}}^{(1)}$ 
accumulates with $n$. Besides this, panel A(b) shows that, for circular states ($1s,~1p,~1d,~1f,~1g$) 
$C_{I_{\pvec}R_{\pvec}}^{(1)}$ diminishes with progression in $r_c$. But for states having one radial node ($2s,~2p,~2d$) there 
appears a maximum in $C_{I_{\pvec}R_{\pvec}}^{(1)}$. This maximum gets right shifted with increase in $l$. Now, panel A(c) of 
Fig.~S4 implies that, the dependence of $C_{I_{t}R_{t}}^{(1)}$ on $r_c$ changes state-wise. Panels B(a), B(b) promptly portray 
that, $C_{I_{\rvec}R_{\rvec}}^{(2)},~C_{I_{\pvec}R_{\pvec}}^{(2)}$, accelerate and decelerate respectively with growth of $r_c$. 
Further, at a fixed $n$, both  $C_{I_{\rvec}R_{\rvec}}^{(2)}$, $~C_{I_{\pvec}R_{\pvec}}^{(2)}$ enhance with emergence of $l$. 
Finally, panel B(c) of Fig.~S4 in SM displays the trend of $C_{I_{t}R_{t}}^{(2)}$ with improvement of $r_c$. A closer investigation 
of Figs.~4 and S4 conveys that, $C_{I_{\rvec}R_{\rvec}}^{(2)},~C_{I_{\pvec}R_{\pvec}}^{(2)},~C_{I_{t}R_{t}}^{(2)}$ characterizes
CHO better than $C_{I_{\rvec}R_{\rvec}}^{(1)},~C_{I_{\pvec}R_{\pvec}}^{(1)},~C_{I_{t}R_{t}}^{(1)}$. Thus, former three measures 
are given in Tables~IV and S4, at some appropriately chosen $r_c$ for which no direct comparison could be made. It is hoped that, 
this study would be useful in future and inspires further work.

\section{Future and outlook}
Four different complexity measures namely $C_{ES},~C_{IS},~C_{ER},~C_{IR}$ are investigated for some low-lying states of CHO in 
composite $r$ and $p$ spaces, keeping $m$ fixed at zero. We have performed the calculations using a global quantity ($E$) 
and a local quantity ($I$). Both these may be used as measure of $order$ in a system. All these 
results are reported here for the first time. Sensitivity of such measures depends on the nature of the particular quantum system
under investigation. It 
is found that, $C^{(1)}_{ES},~C^{(1)}_{ER}$ offer more explicit explanation than $C^{(2)}_{ES},~C^{(2)}_{ER}$ about a given system. 
On the other side, $C^{(2)}_{IS},~C^{(2)}_{IR}$ infer the behavior of CHO more efficiently compared to $C^{(1)}_{IS},~C^{(1)}_{IR}$. 
Hence, considering the nature of complexity measures, it is worthwhile to determine the appropriate value of $b$.  Accurate results 
for $C^{(1)}_{ES},C^{(1)}_{ER},C^{(2)}_{IS},C^{(2)}_{IR}$ are provided for first eight states of CHO. Further, 
an investigation of all these quantities in the realm of Rydberg states under different kinds of soft confined environment 
may be worthwhile pursuing.

\section{Acknowledgement}
Financial support from DST SERB, New Delhi, India (sanction order: EMR/2014/000838) is gratefully acknowledged. 
NM thanks DST SERB, New Delhi, India, for a National-post-doctoral fellowship (sanction order: PDF/2016/000014/CS).

\end{document}

%%%%%%%%%%%%%%%%%%%%%%%%%%%%%%% SUPPLYMENTARY MATERIAL %%%%%%%%%%%%%%%%%%%%%%%%%%%%%%%%%%%%%%%

\pagebreak
\widetext

\begin{center}
\textbf{\large Supplemental Materials: Various Complexity measures in confined isotropic harmonic oscillator.}
\end{center}

\setcounter{page}{1}
\setcounter{figure}{0}
\setcounter{table}{0}

\makeatletter
\renewcommand{\thefigure}{S\arabic{figure}}
\renewcommand{\thetable}{S\arabic{table}}

\vspace{1mm}

\begingroup           %table-S1 
\squeezetable
\begin{table}[h]
\caption{$C_{E_{\rvec}S_{\rvec}}^{(1)},~C_{E_{\pvec}S_{\pvec}}^{(1)}$ and $C_{E_{t}S_{t}}^{(2)}$ for $2p,~1f,~2d,~1g$ 
states in CHA at various $r_c$.}
\centering
\begin{ruledtabular}
\begin{tabular}{l|lll|lll}
$r_c$  &  $C_{E_{\rvec}S_{\rvec}}^{(1)}$ & $C_{E_{\pvec}S_{\pvec}}^{(1)}$  & $C_{E_{t}S_{t}}^{(1)}$\hspace{5mm} & 
$C_{E_{\rvec}S_{\rvec}}^{(1)}$ & $C_{E_{\pvec}S_{\pvec}}^{(1)}$  & $C_{E_{t}S_{t}}^{(1)}$ \vspace{1mm}  \\ 
\hline
\multicolumn{4}{c}{$2p$}  \vline  &   \multicolumn{3}{c}{\hspace{-5mm}$1f$}    \\
\hline
0.1  & 17.254949  & 0.01082 & 0.01082 & 13.123129  & 0.01427    & 0.18738  \\
0.2  & 8.6274798  & 0.02167 & 0.02167 & 6.5615707  & 0.02855    & 0.18738  \\
0.5  & 3.4510731  & 0.05417 & 0.05417 & 2.6247241  & 0.07139    & 0.18738  \\
0.8  & 2.1572081  & 0.08663 & 0.08663 & 1.6407953  & 0.11417    & 0.18733  \\
1.0  & 1.7261513  & 0.10817 & 0.10817 & 1.3131066  & 0.14261    & 0.18726  \\
2.5  & 0.6957840  & 0.25900 & 0.25900 & 0.5416028  & 0.34020    & 0.18425  \\
5.0  & 0.5147727  & 0.51412 & 0.51412 & 0.4457094  & 0.44582    & 0.19870  \\
7.0  & 0.5147878061 & 0.5147737557 & 0.2649992523 &  0.445715446  & 0.445715446 & 0.1986622588   \\
\hline
\multicolumn{4}{c}{$2d$}  \vline  &   \multicolumn{3}{c}{\hspace{-5mm}$1g$}    \\
\hline
0.1  & 16.873312   & 0.01093 & 0.18443 & 13.851781      & 0.01323    & 0.18332    \\
0.2  & 8.4366593   & 0.02186 & 0.18443 & 6.9258936      & 0.02646    & 0.18332    \\
0.5  & 3.3747169   & 0.05466 & 0.18449 & 2.7704060      & 0.06617    & 0.18333    \\
0.8  & 2.1093871   & 0.08742 & 0.18441 & 1.7316772      & 0.10584    & 0.18328    \\
1.0  & 1.6877640   & 0.10920 & 0.18430 & 1.3855802      & 0.13222    & 0.18321    \\
2.5  & 0.6790658   & 0.25869 & 0.17567 & 0.5629633      & 0.31866    & 0.17939    \\
5.0  & 0.4658442   & 0.46532 & 0.21676 & 0.4323887      & 0.43266    & 0.18708    \\
7.0  & 0.4658988655   & 0.4658839758	& 0.2170548158 & 0.4324104301	& 0.4324104287	& 0.1869787794    \\
\end{tabular}
\end{ruledtabular}
\end{table}
\endgroup  

\begin{figure}[h]            %%fig-S1
\centering
\begin{minipage}[c]{0.35\textwidth}\centering
\includegraphics[scale=0.52]{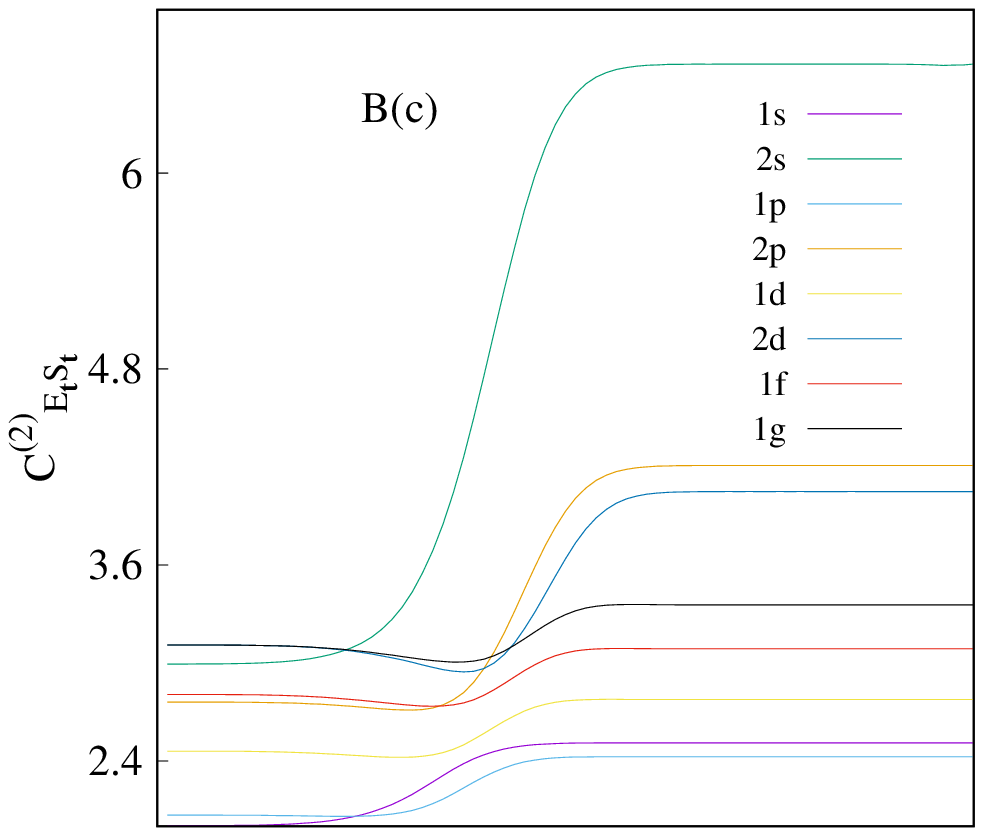}
\end{minipage}%
\vspace{3mm}
\hspace{5in}
\begin{minipage}[c]{0.35\textwidth}\centering
\includegraphics[scale=0.56]{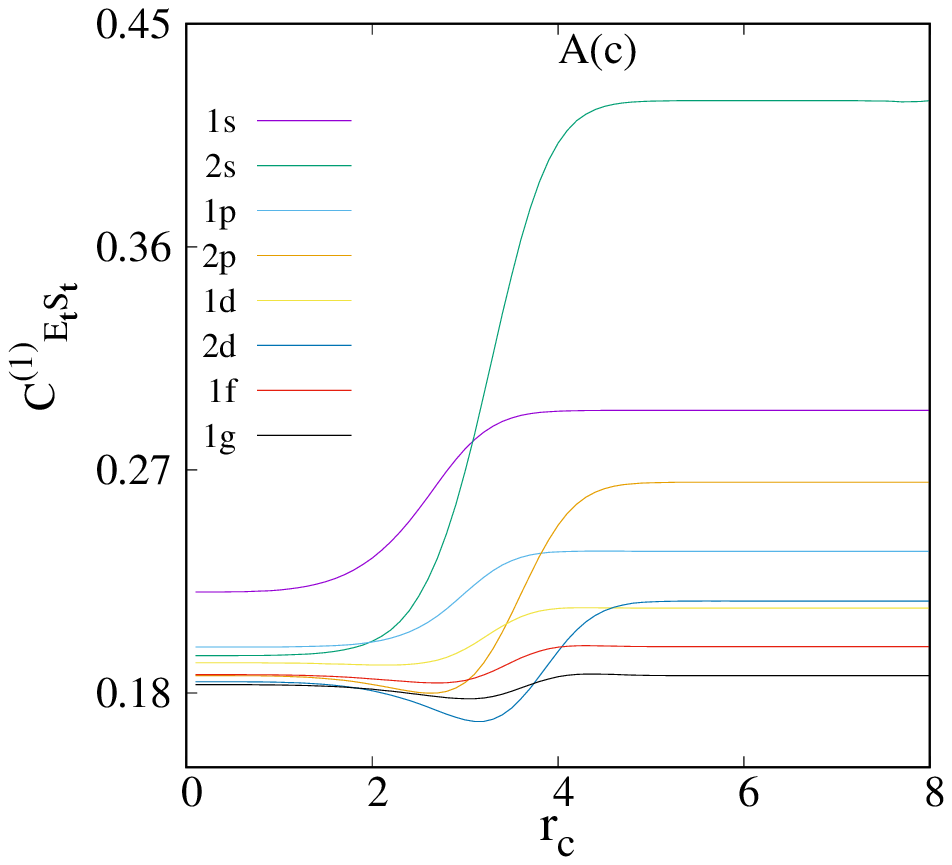}
\end{minipage}%
\caption{Variation of $C_{E_{t}S_{t}}^{(1)}$ (bottom panel) and $C_{E_{t}S_{t}}^{(2)}$ (top panel) in CHA with $r_c$, for $1s, 1p, 1d, 2s, 1f, 2p, 1g, 2d$ states.
 See text for details.}
\end{figure}

\begingroup           %table-S2
\squeezetable
\begin{table}[h]
\caption{$C_{E_{\rvec}R_{\rvec}}^{(1)},~C_{E_{\pvec}R_{\pvec}}^{(1)}$ and $C_{E_{t}R_{t}}^{(1)}$ for $2p,~1f,~2d,~1g$ 
states in CHA at various $r_c$.}
\centering
\begin{ruledtabular}
\begin{tabular}{l|lll|lll}
$r_c$  &  $C_{E_{\rvec}R_{\rvec}}^{(1)}$ & $C_{E_{\pvec}R_{\pvec}}^{(1)}$  & $C_{E_{t}R_{t}}^{(1)}$\hspace{5mm} & 
$C_{E_{\rvec}R_{\rvec}}^{(1)}$ & $C_{E_{\pvec}R_{\pvec}}^{(1)}$  & $C_{E_{t}R_{t}}^{(1)}$ \vspace{1mm}  \\ 
\hline
\multicolumn{4}{c}{$2p$}  \vline  &   \multicolumn{3}{c}{\hspace{-5mm}$1f$}    \\
\hline
0.1  & 21.962364 & 0.00750    & 0.16484    & 15.469104   & 0.00778     &  0.12044    \\
0.2  & 10.981186 & 0.01501    & 0.16484    & 7.7345610   & 0.01557     &  0.12044    \\
0.5  & 4.3925400 & 0.03754    & 0.16490    & 3.0939653   & 0.03893     &  0.12045    \\
0.8  & 2.7455678 & 0.06006    & 0.16489    & 1.9342326   & 0.06228     &  0.12047    \\
1.0  & 2.1967578 & 0.07505    & 0.16488    & 1.5480782   & 0.07784     &  0.12050    \\
2.5  & 0.8799189 & 0.18327    & 0.16126    & 0.6434298   & 0.19071     &  0.12271    \\
5.0  & 0.6848110 & 0.23658    & 0.16201    & 0.5730916   & 0.24199     &  0.13868    \\
7.0  & 0.6852179489   & 0.236635087 & 0.1621466089 & 0.5732802278 & 0.2419896511  & 0.1387278823 \\
\hline
\multicolumn{4}{c}{$2d$}  \vline  &   \multicolumn{3}{c}{\hspace{-5mm}$1g$}    \\
\hline
0.1  & 21.252468 & 0.00653     & 0.13879    & 16.403268  & 0.00670    & 0.11001    \\
0.2  & 10.626237 & 0.01306     & 0.13880    & 8.2016383  & 0.01341    & 0.11001    \\
0.5  & 4.2505534 & 0.03265     & 0.13880    & 3.2807273  & 0.03353    & 0.11001    \\
0.8  & 2.6568022 & 0.05224     & 0.13881    & 2.0507122  & 0.05364    & 0.11001    \\
1.0  & 2.1257180 & 0.06531     & 0.13883    & 1.6409239  & 0.06705    & 0.11002    \\
2.5  & 0.8533617 & 0.16032     & 0.13681    & 0.6693462  & 0.16545    & 0.11074    \\
5.0  & 0.6135813 & 0.21513     & 0.13200    & 0.5534133  & 0.21974    & 0.12160    \\
7.0  & 0.6145939107	& 0.2152218204	& 0.1322740202 & 0.5538899983	& 0.2197225423	& 0.1217021186 \\
\end{tabular}
\end{ruledtabular}
\end{table}
\endgroup 

\begin{figure}[h]            %%fig-S2
\centering
\begin{minipage}[c]{0.35\textwidth}\centering
\includegraphics[scale=0.52]{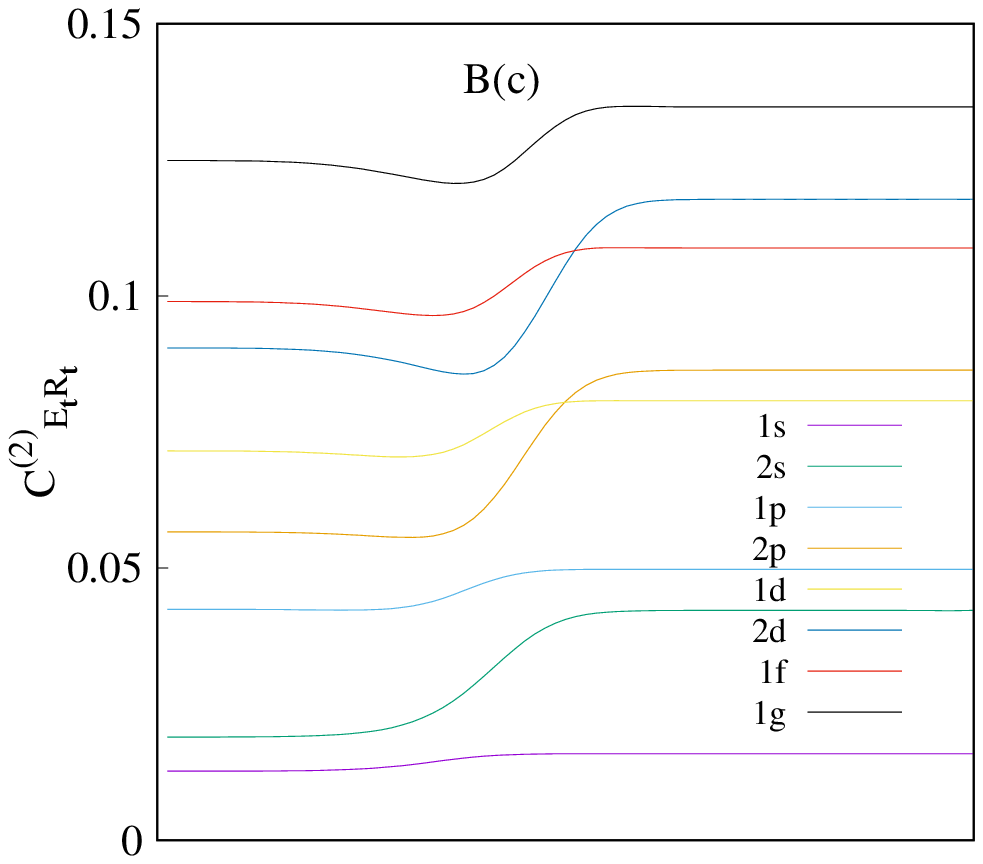}
\end{minipage}%
\vspace{3mm}
\hspace{5in}
\begin{minipage}[c]{0.35\textwidth}\centering
\includegraphics[scale=0.56]{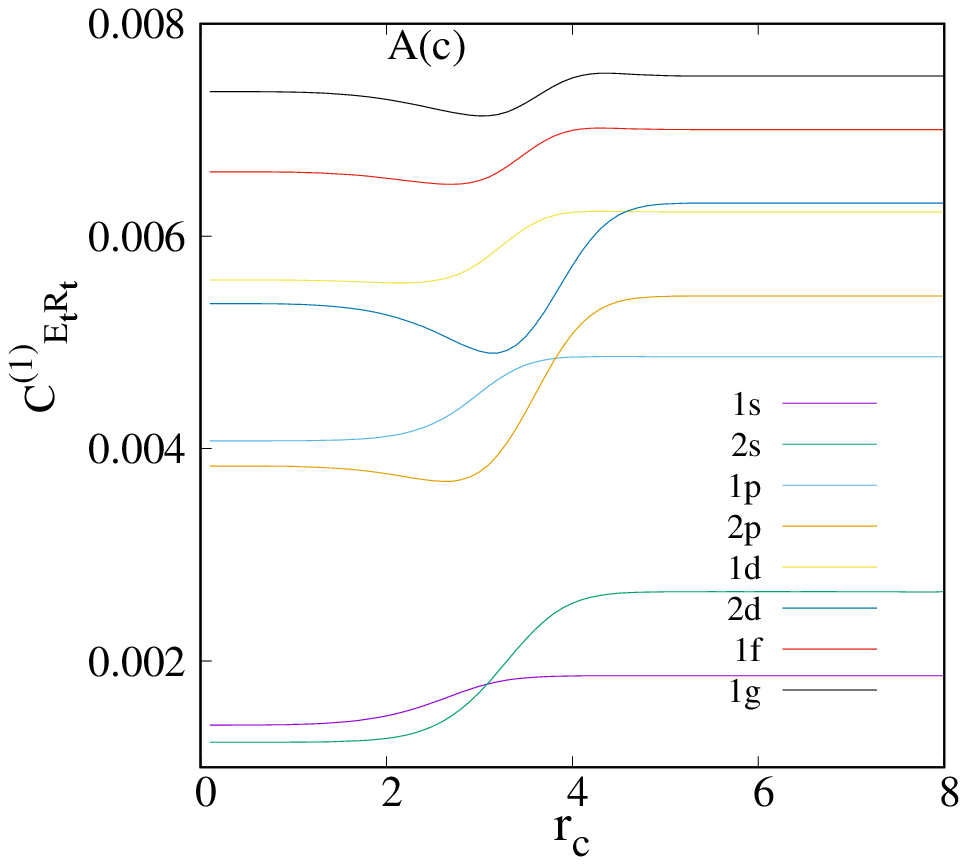}
\end{minipage}%
\caption{Variation of $C_{E_{t}R_{t}}^{(1)}$ (bottom panel) and $C_{E_{t}R_{t}}^{(2)}$ (top panel) in CHA with $r_c$, for $1s, 1p, 1d, 2s, 1f, 2p, 1g, 2d$ states.
 See text for details.}
\end{figure}

\begingroup           %table S3
\squeezetable
\begin{table}[h]
\caption{$C_{I_{\rvec}S_{\rvec}}^{(2)},~C_{I_{\pvec}S_{\pvec}}^{(2)}$ and $C_{I_{t}S_{t}}^{(2)}$ for $2p,~1f,~2d,~1g$ 
states in CHA at various $r_c$.}
\centering
\begin{ruledtabular}
\begin{tabular}{l|lll|lll}
$r_c$  &  $C_{I_{\rvec}S_{\rvec}}^{(2)}$ & $C_{I_{\pvec}S_{\pvec}}^{(2)}$  & $C_{I_{t}S_{t}}^{(2)}$\hspace{5mm} & 
$C_{I_{\rvec}S_{\rvec}}^{(2)}$ & $C_{I_{\pvec}S_{\pvec}}^{(2)}$  & $C_{I_{t}S_{t}}^{(2)}$ \vspace{1mm}  \\ 
\hline
\multicolumn{4}{c}{$2p$}  \vline  &   \multicolumn{3}{c}{\hspace{-5mm}$1f$}    \\
\hline
0.1  & 30.829956   & 34610.5200  & 1067040.8207   & 34.277127        & 37301.84939  & 1278600.25861   \\
0.2  & 61.659980   & 17305.3029  & 1067044.6459   & 68.554152        & 18650.87667  & 1278595.04195    \\
0.5  & 154.156885  & 6922.8172   & 1067199.9473   & 171.374909       & 7459.57109   &  1278383.32131   \\
0.8  & 246.714192  & 4329.2398   & 1068084.9095   & 274.104430       & 4659.46290   &  1277179.42905   \\
1.0  & 308.526800  & 3466.7631   & 1069589.3417   & 342.427973       & 3723.82715   &  1275142.58518    \\
2.5  & 787.931963  & 1469.5392   & 1157896.9377   & 825.488912       & 1420.15678   &  1172323.68189   \\
5.0  & 1139.307518 & 1139.8799   & 1298673.8182   & 1102.110821      & 1101.94896   &  1214469.87865    \\
7.0  & 1139.373375 & 1139.3700   & 1298167.8422   & 1102.13885453152 & 1102.138856  & 1214710.056286      \\
\hline
\multicolumn{4}{c}{$2d$}  \vline  &   \multicolumn{3}{c}{\hspace{-5mm}$1g$}    \\
\hline
0.1  & 45.282593    & 52555.6478   & 2379856.0599   & 46.765278       & 58691.73842    & 2744735.48228    \\
0.2  & 90.565221    & 26277.8642   & 2379860.5853   & 93.530491       &  29345.80827   & 2744727.86739    \\
0.5  & 226.416428   & 10511.8007   & 2380044.3842   & 233.819602      &  11737.33406   & 2744418.78878    \\
0.8  & 362.297120   & 6572.2143    & 2381094.3330   & 374.050977      &  7332.31702    & 2742660.35412    \\
1.0  & 452.937222   & 5260.9704    & 2382889.3444   & 467.435489      &  5861.09040    & 2739681.66713    \\
2.5  & 1142.740818  & 2202.1192    & 2516451.5223   & 1148.845130     &  2247.53126    & 2582065.35261    \\
5.0  & 1772.010616  & 1774.2077    & 3143915.0074   & 1671.968506     &  1671.30118    & 2794362.95082    \\
7.0  & 1772.5172372 & 1772.54450   & 3141865.67995  & 1672.1632856887 &  1672.16328864 & 2796130.05894028 \\
\end{tabular}
\end{ruledtabular}
\end{table}
\endgroup  

\begin{figure}[h]            %%fig-S3
\centering
\begin{minipage}[c]{0.35\textwidth}\centering
\includegraphics[scale=0.52]{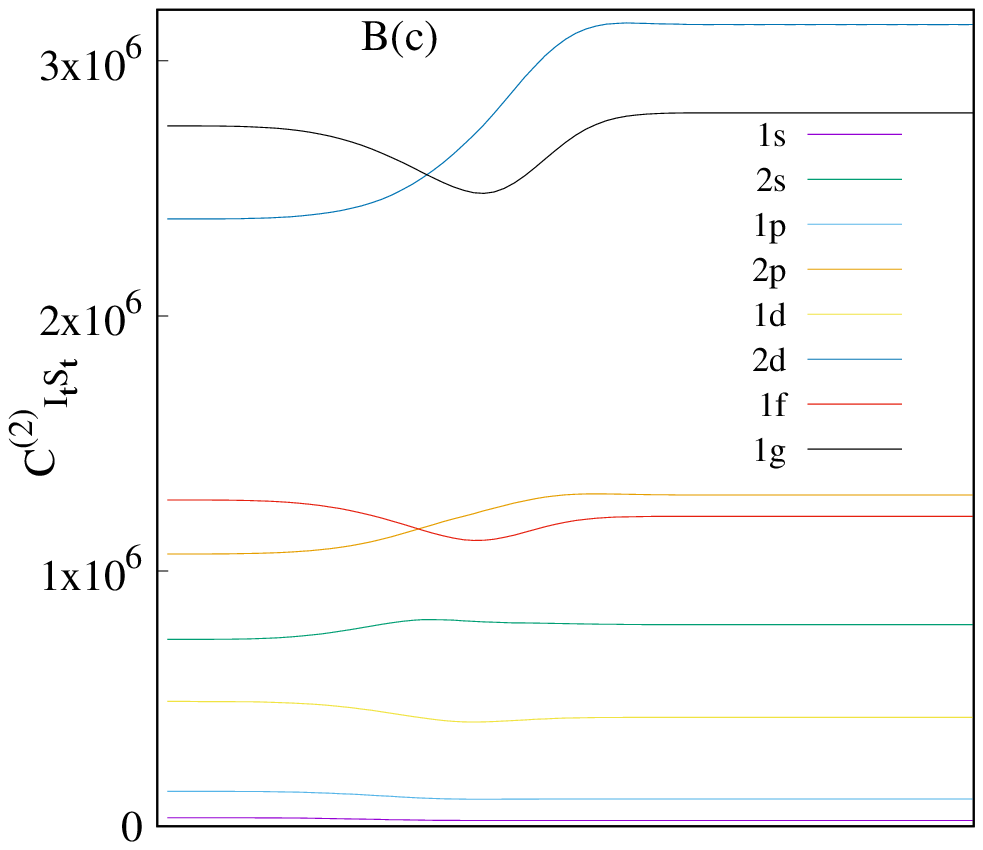}
\end{minipage}%
\vspace{3mm}
\hspace{5in}
\begin{minipage}[c]{0.35\textwidth}\centering
\includegraphics[scale=0.56]{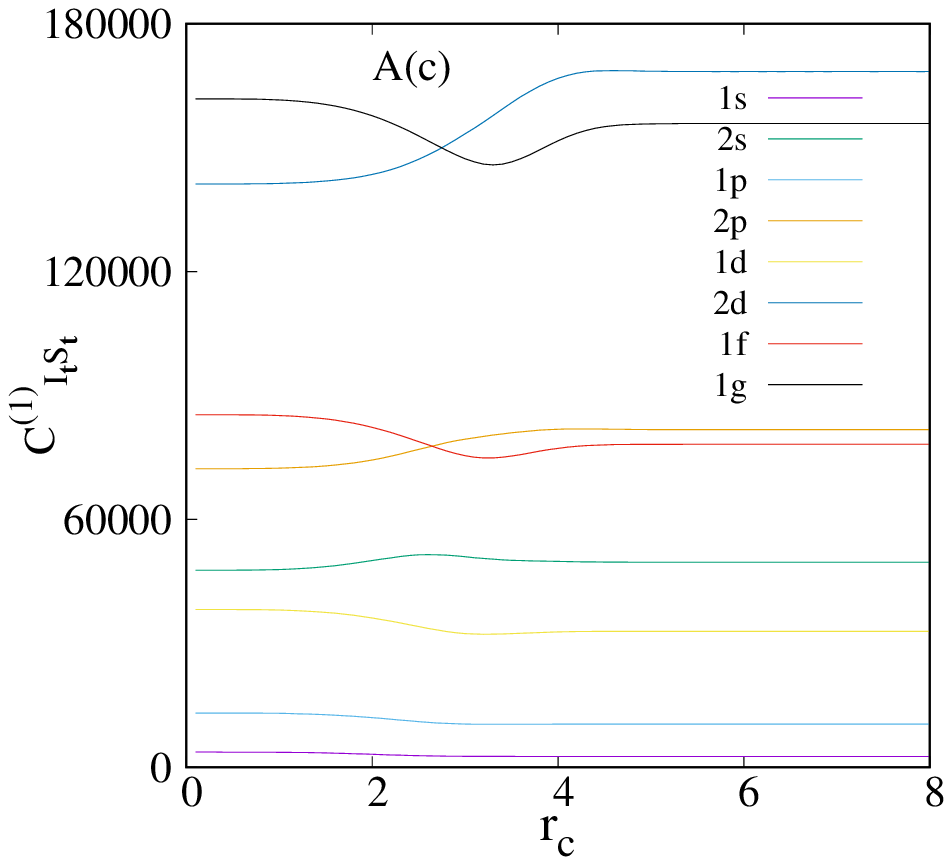}
\end{minipage}%
\caption{Variation of $C_{I_{t}S_{t}}^{(1)}$ (bottom panel) and $C_{I_{t}S_{t}}^{(2)}$ (top panel) in CHA with $r_c$, for $1s, 1p, 1d, 2s, 1f, 2p, 1g, 2d$ states.
 See text for details.}
\end{figure}

\begingroup           %table S4
\squeezetable
\begin{table}[h]
\caption{$C_{I_{\rvec}R_{\rvec}}^{(2)},~C_{I_{\pvec}R_{\pvec}}^{(2)}$ and $C_{I_{t}R_{t}}^{(2)}$ for $2p,~1f,~2d,~1g$ 
states in CHA at various $r_c$.}
\centering
\begin{ruledtabular}
\begin{tabular}{l|lll|lll}
$r_c$  &  $C_{I_{\rvec}R_{\rvec}}^{(2)}$ & $C_{I_{\pvec}R_{\pvec}}^{(2)}$  & $C_{I_{t}R_{t}}^{(2)}$\hspace{5mm} & 
$C_{I_{\rvec}R_{\rvec}}^{(2)}$ & $C_{I_{\pvec}R_{\pvec}}^{(2)}$  & $C_{I_{t}R_{t}}^{(2)}$ \vspace{1mm}  \\ 
\hline
\multicolumn{4}{c}{$2p$}  \vline  &   \multicolumn{3}{c}{\hspace{-5mm}$1f$}    \\
\hline
0.1  & 44.271169   & 19970.5174 & 884118.1701   & 43.867820      & 15021.73056 & 658970.57651   \\
0.2  & 88.542409   & 9985.3085  & 884123.2806   & 87.735537      & 7510.86091  & 658969.42157   \\
0.5  & 221.363121  & 3994.9328  & 884330.7969   & 219.328424     & 3004.27340  & 658922.55152   \\
0.8  & 354.245464  & 2499.7145  & 885512.5259   & 350.830623     & 1877.41927  & 658656.17667   \\
1.0  & 442.942959  & 2003.6869  & 887519.0282   & 438.337433     & 1501.59674  & 658206.06330   \\
2.5  & 1120.574981 & 874.6934   & 980159.6435   & 1068.910611    & 596.06551   & 637140.75398   \\
5.0  & 1748.130191 & 355.5763   & 621593.7685   & 1606.880150    & 440.68957   & 708135.33589   \\
7.0  & 1749.712868 & 355.093236 & 621311.205169 & 1607.68230133  & 440.9049523 & 708835.0883814 \\
\hline
\multicolumn{4}{c}{$2d$}  \vline  &   \multicolumn{3}{c}{\hspace{-5mm}$1g$}    \\
\hline
0.1  & 64.009563   & 24273.2716 & 1553721.5208   & 60.264374      & 21170.66473 & 1275836.85768 \\
0.2  & 128.019163  & 12136.6807 & 1553727.7205   & 120.528683     & 10585.32761 & 1275835.60452 \\
0.5  & 320.051676  & 4855.4017  & 1553979.4627   & 301.315183     & 4234.05396  & 1275784.74865 \\
0.8  & 512.117124  & 3037.2289  & 1555416.9651   & 482.044895     & 2646.01023  & 1275495.72562 \\
1.0  & 640.219512  & 2433.3390  & 1557871.1491   & 602.430376     & 2116.43936  & 1275007.36035 \\
2.5  & 1609.829117 & 1074.3700  & 1729552.1569   & 1489.421348    & 840.86097   & 1252396.28603 \\
5.0  & 2678.634631 & 557.7446   & 1493994.0015   & 2420.979640    & 604.92919   & 1464521.26703 \\
7.0  & 2685.563139 & 556.522132 & 1494575.284942 & 2424.207149829 & 605.68437   & 1468304.38029 \\
\end{tabular}
\end{ruledtabular}
\end{table}
\endgroup  
 
\begin{figure}[h]        %%fig-S4
\centering
\begin{minipage}[c]{0.35\textwidth}\centering
\includegraphics[scale=0.52]{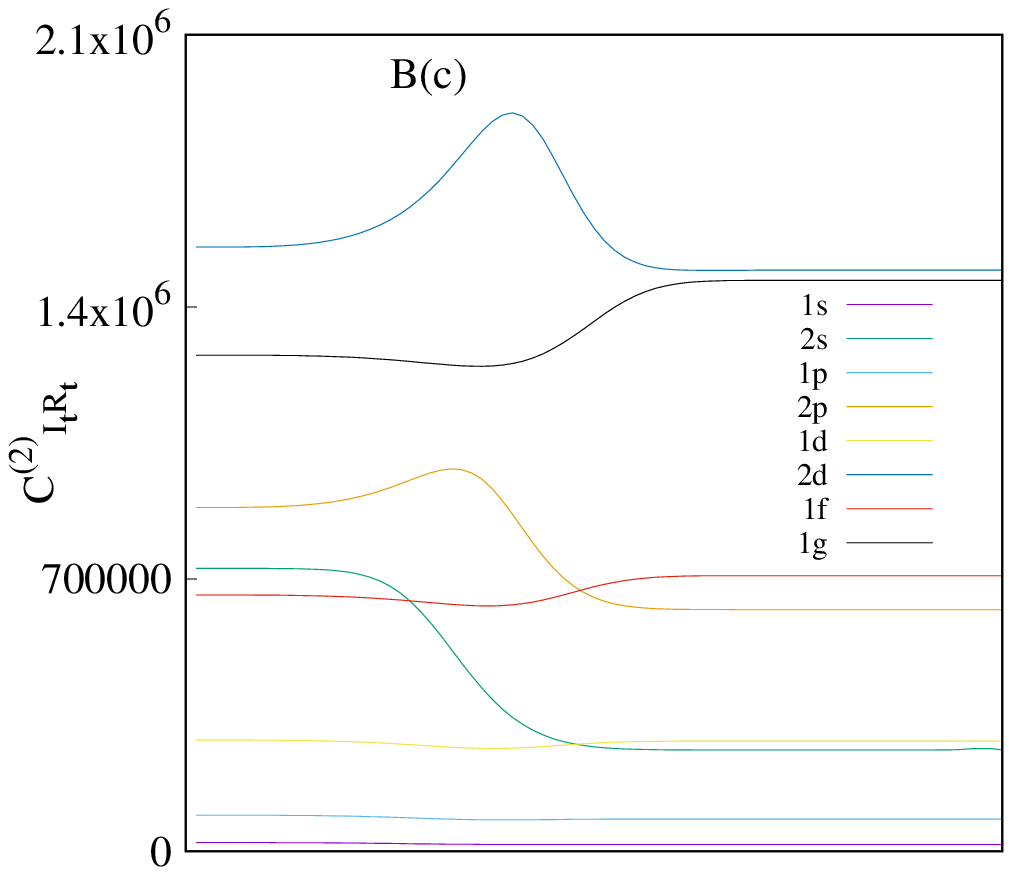}
\end{minipage}%
\vspace{3mm}
\hspace{5in}
\begin{minipage}[c]{0.35\textwidth}\centering
\includegraphics[scale=0.56]{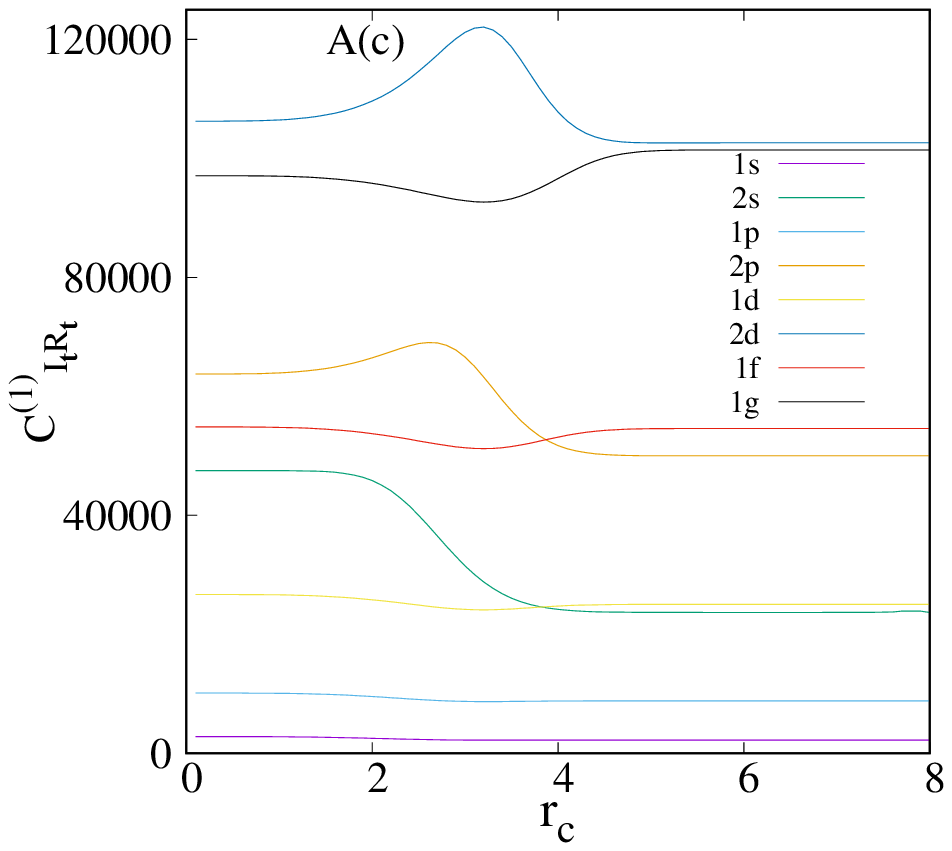}
\end{minipage}%
\caption{Variation of $C_{I_{t}R_{t}}^{(1)}$ (bottom panel) and $C_{I_{t}R_{t}}^{(2)}$ (top panel) in CHA with $r_c$, for $1s, 1p, 1d, 2s, 1f, 2p, 1g, 2d$ states.
 See text for details.}
\end{figure}